\documentclass[lettersize,journal]{IEEEtran}
\usepackage{amsmath,amsfonts}
\usepackage{amsthm}
\usepackage{xcolor}
\usepackage{algorithm}
\usepackage{algpseudocode}
\usepackage{array}
\usepackage{caption}
\usepackage[caption=false,font=normalsize,labelfont=sf,textfont=sf]{subfig}
\usepackage{textcomp}
\usepackage{stfloats}
\usepackage{url}
\usepackage{verbatim}
\usepackage{graphicx}
\usepackage{cite}
\usepackage{amssymb}
\usepackage{bbold}
\DeclareMathOperator{\E}{\mathbb{E}}
\newcommand{\probP}{\text{I\kern-0.15em P}}
\DeclareMathOperator{\e}{e}

\theoremstyle{remark}

\newtheorem*{example}{Example}

\theoremstyle{definition}
\newtheorem{definition}{Definition}[section]

\theoremstyle{plain}

\newtheorem{setting}{Setting}

\begin{document}

\title{Bidding efficiently in Simultaneous Ascending Auctions with incomplete information using Monte Carlo Tree Search and determinization}

\author{Alexandre Pacaud, Aurelien Bechler and Marceau Coupechoux
        % <-this % stops a space
\thanks{A.Pacaud and A.Bechler are with Orange Labs, France (e-mail: alexandre.pacaud@orange.com, aurelien.bechler@orange.com).}
\thanks{A.Pacaud and M.Coupechoux are with LTCI, Telecom Paris, Institut Polytechnique de Paris, France (e-mail: marceau.coupechoux@telecom-paris.fr). The work of M. Coupechoux has been performed at LINCS (lincs.fr).}
}

% The paper headers
\markboth{Bidding efficiently in Simultaneous Ascending Auctions with incomplete information}%
{Shell \MakeLowercase{\textit{et al.}}: A Sample Article Using IEEEtran.cls for IEEE Journals}

\maketitle

\begin{abstract}
For decades, Simultaneous Ascending Auction (SAA) has been the most widely used mechanism for spectrum auctions, and it has recently gained popularity for allocating 5G licenses in many countries. Despite its relatively simple rules, SAA introduces a complex strategic game with an unknown optimal bidding strategy. Given the high stakes involved, with billions of euros sometimes on the line, developing an efficient bidding strategy is of utmost importance. In this work, we extend our previous method, a Simultaneous Move Monte-Carlo Tree Search (SM-MCTS) based algorithm named $SMS^{\alpha}$ to incomplete information framework. For this purpose, we compare three determinization approaches which allow us to rely on complete information SM-MCTS. This algorithm addresses, in incomplete framework, the four key strategic issues of SAA: the exposure problem, the own price effect, budget constraints, and the eligibility management problem. Through extensive numerical experiments on instances of realistic size with an uncertain framework, we show that $SMS^{\alpha}$ largely outperforms state-of-the-art algorithms by achieving higher expected utility while taking less risks, no matter which determinization method is chosen.

\end{abstract}


\section{Introduction}

For decades, mobile operators have competed for cellular spectrum to improve service quality and develop wireless communication networks\footnote{Part of the material exposed in this paper has been presented in the PhD thesis of the first author~\cite{pacaud202thesis}.}. Access to a wide range of frequencies is essential for optimal performance. For instance, while low-frequency waves offer longer transmission distances and better obstacle penetration compared to high-frequency waves, they usually provide lower data rates. Therefore, both frequency ranges complement each other. Today, cellular spectrum is mostly allocated in the form of licences through auctions. Over the past 20 years, the preferred auction mechanism for the sale of spectrum licences has been the Simultaneous Ascending Auction (SAA), also known as \textit{Simultaneous Multi Round Auction} (SMRA). It features a dynamic multi-round format where bidders simultaneously submit bids for all licenses, ending when no new bids are made in a round. SAA's popularity stems from its relatively simple rules and substantial revenue generation for regulators. It has recently been used in Germany \cite{5G_germany}, Portugal \cite{5G_portugal}, and Italy \cite{5G_italy} for the sale of 5G licences and is anticipated to play a central role in 6G frequency allocation by 2030. Both of its creators, Paul Milgrom and Robert Wilson, were jointly awarded the 2020 Sveriges Riksbank Prize in Economic Sciences in Memory of Alfred Nobel, mainly for their contributions to SAA.
Given the immense financial stakes, sometimes reaching billions of euros in SAA, and the pivotal role of acquired licenses in an operator's business plan, having an efficient bidding strategy is crucial. However, traditional auction theory and exact game resolution methods often fail to compute the optimal bidding strategy due to the inherent complexity of the bidding game \cite{reeves2005exploring}. For instance, SAA induces a $n$-player simultaneous move game with incomplete information with a large state and action space. Additionally, various complex strategic issues arise from the mechanism of SAA, as well as the specific constraints faced by bidders. 

The four main strategic issues of SAA are the \textit{exposure problem}, the \textit{own price effect}, \textit{budget constraints} and the \textit{eligibility management problem}. The exposure problem occurs when a bidder competes for a set of complementary licenses and ends up paying excessively for the ones acquired due to unexpectedly fierce competition. The own price effect refers to the decrease in bidders' utility as prices escalate. Therefore, refraining from bidding on certain licenses to keep prices low can be an efficient bidding strategy. Budget constraints refer to the limited funds allocated to bidders for the entire auction. Inadequate allocation of these funds towards the desired licenses can result in suboptimal outcomes. The eligibility management problem arises from activity rules that penalise bidders failing to maintain a minimum level of bidding activity. It introduces a quantity named \textit{eligibility}, dictating the items a bidder can bid on. Violating activity rules reduces a bidder's eligibility, restricting them from bidding on certain item sets for the remainder of the auction. Efficiently managing eligibility is crucial to avoid being trapped in a disadvantaged position and secure a favourable outcome. In this work, we propose new bidding strategies that addresses these challenges in an incomplete information framework.

\subsection{Related work}

Many studies on SAA, such as \cite{milgrom2000putting,cramton2002spectrum,cramton2006simultaneous}, focus on its mechanism design, efficiency, and revenue generation for the regulator. However, only a few delve into the strategic bidding problem. These studies often aim to tackle one of SAA's main strategic issues within specific contexts and simplified versions of the auction. Additionally, the proposed solutions are often applicable only to small instances. Below, we introduce some relevant works that consider the bidder's perspective.

The exposure problem is considered as the most problematic strategic issue in SAA. To simplify the problem, some works have considered a clock-format of SAA \cite{goeree2014equilibrium} with only two types of bidders: \textit{local} and \textit{global}. This format offers the advantage of being a tractable model under certain conditions, enabling bidders to have continuous and differentiable expected utilities. Consequently, standard optimisation techniques can be employed to derive an equilibrium. In \cite{goeree2014equilibrium}, Goeree and Lien use this format to obtain a Bayesian framework and compute the global bidder's optimal drop-out level when all items are identical and all bidders have super-additive value functions. They extend their work to the case of two global bidders with regional complementarities. By introducing a pause system in the clock-format format of SAA that enables jump bidding, Zheng \cite{zheng2012jump} fully eliminates the exposure problem for instances with two items and one global bidder by building a continuation equilibrium. In the original format of SAA \cite{cramton2006simultaneous}, Wellman et al. \cite{wellman2008bidding} introduce a bidding strategy addressing the exposure problem by predicting closing prices. While initial results appeared promising, they were only obtained for particular super-additive value functions.
Regarding own price effect, Milgrom \cite{milgrom2000putting} describes a collusive equilibria in an SAA with complete information, two items and two bidders having additive value functions. This work was then extended to incomplete information and super-additive value functions by Brusco and Lopomo \cite{brusco2002collusion} by constructing a collusive equilibria through signalling. Similarly to their algorithm tackling the exposure problem, Wellman et al. \cite{wellman2008bidding} propose a bidding strategy based on a prediction of closing prices to address the own price effect when all items are identical and bidders possess subadditive value functions. Despite their efforts, the results proved disappointing, as its performance is significantly outperformed by a basic demand reduction approach. In a clock-format within a specific context involving many constraints such as complete information and identical goods, Riedel and Wolfstetter \cite{riedel2006immediate} prove that the only Nash equilibrium surviving iterated elimination of dominated strategies is the strategy profile where bidders reduce their demand in the initial round to achieve efficient allocation.
Using a specific clock-format with a pause system, Brusco and Lopomo \cite{brusco2009simultaneous} analyse various inefficiencies arising from the interaction of the exposure problem and binding budgets. They focus on the structure of noncollusive equilibria and obtain results within the context of auctions involving two items and two global bidders with super-additive value functions. In \cite{bulow2009winning}, Bulow et al. propose a final price forecasting approach by tracking bidders' \textit{bid exposure} to mitigate the exposure problem in case of binding budgets. 
Regarding the eligibility management problem, little work has been proposed. However, it is commonly accepted that one should gradually decrease its eligibility to avoid being trapped in a disadvantageous position \cite{weber1997making}. Thus, bidders may bid on unwanted items to maintain eligibility and deceive others, a tactic known as \textit{parking} \cite{porter2006fcc}.

To solve the bidding problem in SAA, we have proceeded step by step gradually increasing the complexity of our SAA model. In \cite{pacaud2022monte}, we consider a turn-based deterministic model with perfect and complete information with no activity rule or budget constraints for which we propose a bidding strategy based on Monte Carlo Tree Search (MCTS). In \cite{pacaud2024bidding}, we consider the original format of SAA with complete information for which we propose a bidding strategy computed by Simultaneous Monte Carlo Tree Search (SM-MCTS). 
In this paper, we extend our work to incomplete information. Bidders are now faced with the challenge of not knowing their opponents' value functions and budgets. Incomplete information games are often referred to as Bayesian Games \cite{harsanyi1968games}. In this work, we decide to use a similar Bayesian setting as in \cite{nedelec2022learning} to model the incertitude on one’s value function and budget. Thus, players no longer have a point-wise estimate of the value function or budget of their opponents, as it was the case for complete information, but a probability distribution. % A widely made assumption is that these
%distributions are common knowledge between bidders \cite{nedelec2022learning}.}
In the literature, we identify three types of MCTS which are applied to imperfect information games: \textit{Determinized MCTS} \cite{whitehouse2011determinization}, \textit{Information-State MCTS} \cite{cowling2012information} and \textit{Belief-State MCTS} \cite{wang2015belief}. As shown later in this paper, methods for imperfect information games can easily be applied to incomplete information games. Only versions of the first two types of MCTS will be studied in this work.

\subsection{Contributions}

In this paper, we consider the original format of SAA with incomplete information for which we propose three efficient bidding strategies tackling simultaneously its four main strategic issues. We make the following contributions: 

\begin{itemize}
    \item We model the auction as an $n$-player simultaneous move game with incomplete information that we name SAA-inc. No specific assumption is made on the bidders' value function. 
    \item We present three efficient bidding strategies that tackle simultaneously the \textit{exposure problem}, the \textit{own price effect}, \textit{budget constraints} and the \textit{eligibility management problem} in SAA-inc. These approaches combine determinization with Simultaneous Move Monte Carlo Tree Search (SM-MCTS) \cite{tak2014monte}. Determinization is built upon a variant of our complete information solution~\cite{pacaud2024bidding} using EXP3 during the selection phase. We also propose an inference method based on tracking bid exposure for updating one’s belief about an opponent’s budget. To the best of our knowledge, these are the first algorithms that tackle the four main strategic issues of SAA with incomplete information. 
    \item In order to compare to state-of-the-art solutions, we propose a general framework for generating value functions and budgets with a certain level of uncertainty. Through extensive experiments, we show that our three determinization approaches significantly outperform state-of-the-art bidding strategies in budget and eligibility constrained environments with incomplete information. Our analysis of performance indicators sheds light on the impact of uncertainty on our determinization approaches, including their effect on coordination.
\end{itemize}

The remainder of this paper is structured as follows. In Section~\ref{SAA}, we define our SAA-inc model, formulate the problem as a bidding game, derive its complexity, and introduce performance indicators. 
In Section~\ref{determinization approaches}, we propose three determinization approaches that extends the algorithm proposed in~\cite{pacaud2024bidding} to incomplete information. A general framework for generating value and budget distributions as well as the selection of relevant samples is presented in Section~\ref{selecting relevant combinations of profiles}. Section~\ref{extensive experiments} is dedicated to numerical results. Section~\ref{sec:conclusion} concludes the paper. 

\section{Simultaneous Ascending Auction} \label{SAA}

\subsection{Mechanism and rules of our SAA model}

The Simultaneous Ascending Auction (SAA) \cite{milgrom2000putting,cramton2006simultaneous,wellman2008bidding} is a widely-used mechanism design for selling $m$ indivisible goods to $n$ players through separate and concurrent English auctions. Bidding unfolds in multiple rounds, during which players submit their bids simultaneously. In each round, the player with the highest bid for an item becomes its temporary winner. If many players submit the same highest bid for an item, the temporary winner is randomly selected among them. The bid price of an item $j$, denoted as $P_j$, is then set to the highest bid placed on it. At the conclusion of each round, the temporary winners and bid prices are disclosed to all players. The auctions closes if no new bids are submitted during a round, and the items are subsequently sold to their corresponding temporary winners at their current bid prices.
To discourage bidders from reducing their bidding activity, we apply the following activity rule, commonly used in the literature due to its simplicity: the number of items $e_i$ for which a player $i$ participates (temporarily wins or bids) can never increase \cite{goeree2014equilibrium, milgrom2004putting,pacaud2024bidding}. 

In our model, all bid prices are initialized to $0$ at the beginning of the auction. To reduce the bidding space, we adopt two simplifications commonly found in the literature. Firstly, we restrict new bids to $P_j+\varepsilon$, where $\varepsilon$ represents a fixed bid increment \cite{goeree2014equilibrium,wellman2008bidding,pacaud2022monte,pacaud2024bidding}. Secondly, we assume that bidders will not bid on items they are currently temporarily winning \cite{wellman2008bidding,pacaud2022monte,pacaud2024bidding}. 

\subsection{Bidders' modelling}

Each bidder $i$ is defined by three quantities: their value function $v_i$, their budget $b_i$ and their eligibility $e_i$. The two first characteristics are private information and the last is public knowledge. Without loss of generality, we assume that $v_i$ and $b_i$ are chosen independently. If the current bid price vector is $P$, a player $i$ temporarily winning a set of items $Y$ with current eligibility $e_i$ can bid on a set of items $X$ if and only if 
\begin{equation}
    \begin{cases}
	|X|+|Y|\leq e_i \hspace*{2.9cm} \text{(eligibility constraint)}\\
    \sum_{j\in X}(P_j+\varepsilon)\leq b_i-\sum_{j\in Y} P_j \quad \text{(budget constraint)}
	\end{cases}
\end{equation}

If player $i$ ends up winning a set of items $X$ at bid price vector $P$, their utility $\sigma_i$ is equivalent to their profit, i.e.: 
\begin{equation}
    \label{profit}
    \sigma_i(X,P)=v_i(X)-\sum_{j\in X} P_j 
\end{equation}
Value functions are assumed to be normalised ($v_i(\emptyset)=0$), finite and verify the free disposal condition, i.e. for any two sets of goods $X$ and $Y$ such that $X\subset Y$, then $v(X)\leq v(Y)$ \cite{lehmann2006combinatorial,milgrom2000putting}.
For the definition of our strategies, we moreover rely on a specific utility function named \textit{risk-averse utility} defined as~\cite{pacaud2024bidding}:
\begin{equation}
    \label{risk-averse utility}
    \sigma_i^\alpha(X,P)=(1+\alpha \mathbb{1}_{\sigma_i(X,P)<0})\sigma_i(X,P)
\end{equation}
Compared to the utility previously defined, losses are more heavily penalised. The hyperparameter $\alpha$ allows for balancing between expected utility and risk aversion.

\subsection{Modelling uncertainty with types and profiles}

To model the uncertainty about the value function and budget of one's opponents, we rely on the notion of \textit{types}. By the term ``type," we denote a possible combination of one’s private information. For example, if player $1$ has two possible value functions, $v_1^1$ and $v_1^2$, as well as two possible budgets, $b_1^1$ and $b_1^2$, then from the perspective of player $2$, player $1$ presents $4$ possible types: ($v_1^1$, $b_1^1$), ($v_1^1$, $b_1^2$), ($v_1^2$, $b_1^1$), and ($v_1^2$, $b_1^2$).

Each player maintains a probability distribution regarding the types of their opponents. A common assumption is that all opponents of player $i$ share the same probability distribution concerning the type of $i$ \cite{nedelec2022learning}. Therefore, we denote $\mathcal{F}_i$ and $\mathcal{B}_i$ as the value distribution and budget distribution of bidder $i$, respectively. Both distributions are independent. We define the type distribution $\mathcal{T}_i$ of player $i$ as the joint probability that $i$ possesses a value function of $v$ and a budget of $b$, i.e.,
\begin{equation}
    \mathcal{T}_i(v,b)=\mathcal{F}_i(v)\mathcal{B}_i(b)
\end{equation}

We define $\mathcal{T}_{-i}$ as the joint distribution of all type distributions of all players except $i$. Each bidder $i$ knows their own type but does not know the types of their opponents, i.e., a bidder $i$ only knows $v_i$, $b_i$ and $\mathcal{T}_{-i}$. We call \textit{SAA-inc} this simplified version of SAA with the above uncertainty and bidders’ modelling. 

Two of our proposed determinization algorithms rely on the sampling of the type distribution. As the type distribution support can be very general, the sampling may be either very time-consuming or very inaccurate when a time budget is imposed. As a consequence, we define and select a smaller set of parameterized {\it profiles} where the value function and the budget are derived from the type distribution, however with a smaller support. The way profiles are selected is explained in Section~\ref{selecting relevant combinations of profiles}.

\subsection{Extensive form}

Multi-round games are typically represented by a finite rooted directed tree known as the extensive form. This game tree features two types of nodes: decision nodes and chance nodes. Decision nodes denote points in the game where players must take decisions, with each possible decision represented by an outgoing directed edge. Chance nodes represent uncertain outcomes in the game, with each outgoing edge associated with a predetermined fixed probability. Information sets are introduced to account for the player's lack of information in determining the exact decision node they are playing. These sets regroup decision nodes that are indistinguishable to the player, based on the available information at this stage of the game.

An incomplete information game, also referred to as a Bayesian game \cite{harsanyi1968games}, can be seen as a game with complete but imperfect information. Indeed, a Bayesian game can be equivalently represented by a game where Nature initially assigns each player's type based on their type distribution, followed by the players playing in the resulting sub-game with complete but imperfect information. This representation is known as the Bayes-equivalent of the original game \cite{harsanyi1968games}. We rely on this representation to model our bidding game. From now on, we implicitly identify the SAA-inc game with its Bayes-equivalent and use both interchangeably. 

In each sub-game, decision nodes represent different game states, while chance nodes denote the random draws of temporary winners in case of identical highest submitted bids. At each decision node, outgoing edges represent the different sets of items a player bids on if they select that particular edge. Each state is defined  by seven features: the identity of the bidder playing at that state (we name it the ``concerned player"), the eligibility vector at the start of the current round, the temporary winner of each item, the current bid price vector, the concerned player's type, the bids already submitted during the current round and the types of the concerned player's opponents. The first five features are common knowledge and the last two are hidden information for the concerned player. Thus, all states which differ only by the last two features belong to the same information set.

\begin{figure}[t]
\centering
\includegraphics[width=\linewidth]{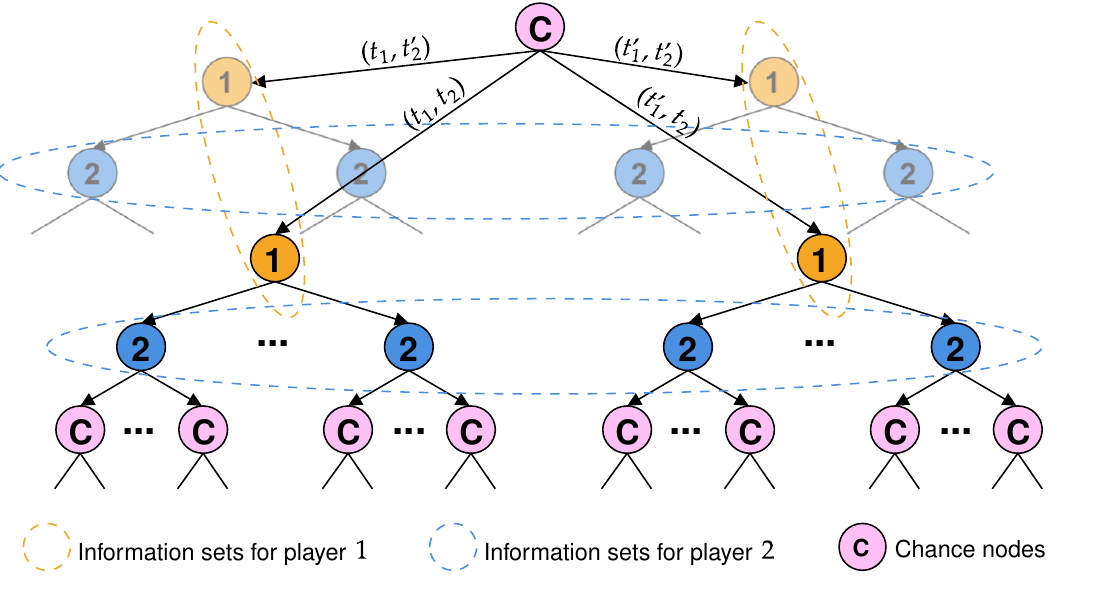}
\caption{Extensive form of a two player SAA-inc game with information sets and chance nodes. The first chance node corresponds to the draw by Nature of each player's type. For example, the couple $(t_1,t'_2)$ means the first player is type $t_1$ and the second player is type $t'_2$.} 
\label{fig:SAA-inc}
\end{figure}

An example of the extensive form of an SAA-inc game between two players is represented in Figure~\ref{fig:SAA-inc}. At the beginning of the game, Nature draws a type $t_i$ for each player $i$. Each player knows their type but not their opponent's type. Hence, each player does not know exactly in which sub-game it is playing and, thus, different states across sub-games can belong to a same information state. For instance, all states corresponding to the first round of the auction which share the same concerned player and the same concerned player's type belong to the same information set. Thus, in Figure~\ref{fig:SAA-inc}, the two nodes corresponding to player $1$ with types $(t_1,t_2)$ and $(t_1,t'_2)$ belong to the same information set. 

It is important to note that we are able to represent the SAA-inc game in extensive form because the players' type distributions are not continuous but discrete. This is because value functions and budgets correspond in practice to monetised values and we only bid on multiple of the price increment. 

\subsection{Game and strategic complexities}

\subsubsection{Game complexities}

To measure the complexity of the SAA-inc game, we focus on two metrics: \textit{information set space complexity} \cite{pacaud2024bidding} and \textit{game tree complexity} \cite{van2002games}. The first refers to the number of different information sets in the game. Given that a strategy is a function that assigns an action to each information set, the information set space complexity corresponds to the size of the domain of a strategy. The second corresponds to the number of different paths in its game tree. Lower bounds can be computed by assuming unlimited budgets and ignoring activity rules.

In this case, we can show that the number of possible information sets of a given instance of the SAA-inc game is given by $\sum_{i=1}^n |supp(\mathcal{T}_i)| (Rn+1)^m$, where $|supp(\mathcal{T}_i)|$ is the cardinal of the support of the discrete type distribution of player $i$ and $R$ is the number of rounds. A lower bound for the game tree complexity is $\Omega(2^{m(n-1)R}\prod_{i=1}^n |supp(\mathcal{T}_i)|)$. These two bounds show the very high complexity of the SAA-inc game, see~\cite{pacaud2024bidding} for a numerical example in the complete information case.

\subsubsection{Strategic complexities}

In addition to the high game complexity, the SAA-inc game presents also complex strategic issues. We present hereafter the four main ones, see \cite{pacaud2024bidding} for numerical simple examples.

\textbf{Exposure:} It is an issue that arises when a player attempts to acquire complementary items but ends up paying too much for those obtained, resulting in negative utility.

\textbf{Own price effect:} This issue arises because each bid on an item increases its price, consequently reducing the utility of all players interested in acquiring it. Thus, bidders have a mutual interest in keeping prices low. To address this problem, a bidder $i$ may choose to stop bidding on certain items, in the hope that their opponents will refrain from bidding on the items that $i$ is temporarily winning, a strategy known as demand reduction \cite{weber1997making,ausubel2014demand}. Alternatively, players may implicitly coordinate to divide the items among themselves, referred to as collusion \cite{brusco2002collusion}. As explicit communication is prohibited, this implicit coordination relies heavily on the accuracy of the estimates of one's opponents' types.

\textbf{Budget constraints:} Setting a maximum limit on the amount a bidder can spend during an auction can significantly influence its outcome. This restriction can deter players from bidding on certain sets of items and highly affect the risk of exposure. Furthermore, armed with this knowledge, players may substantially alter their bidding strategies.

\textbf{Eligibility management:} Effectively managing one's eligibility is crucial for achieving a favourable outcome. While bidding on numerous items to maintain a high eligibility may seem advantageous at first glance, it triggers the own price effect, making it often strategically suboptimal. Conversely, reducing eligibility to facilitate collusions can leave a bidder vulnerable if opponents deviate from expected behaviour. Thus, finding the right balance is essential.

\subsection{Performance indicators} \label{perf indicators}

As in \cite{pacaud2024bidding}, we consider the following performance metrics for a strategy: the \textit{expected utility}, the \textit{expected exposure}, the \textit{exposure frequency}, the \textit{average price paid per item won} and the \textit{ratio of items won}. The expected exposure is the opposite of the sum of losses incurred by a strategy divided by the number of times it has been used. The exposure frequency is the number of times a strategy has been exposed divided by the number of times it has been used. The average price paid per item won allows us to highlight the increase in expected utility resulting from a better management of the own price effect. The ratio of items won allows us to see whether a strategy effectively divides items among players and that no item remains unnecessarily unsold.

\section{Three determinization approaches for incomplete information} \label{determinization approaches}

\subsection{Bidding strategy for complete information SAA} 

In \cite{pacaud2024bidding}, we have developed an efficient bidding strategy, named $SMS^\alpha$, for SAA-inc in the specific case of complete information (SAA-c) based on a tuned Simultaneous Move Monte Carlo Tree Search (SM-MCTS) \cite{tak2014monte, bovsansky2016algorithms}. The four classical MCTS phases, i.e. selection, expansion, rollout and backpropagation have been modified and adapted, with the use of a specific risk-averse utility function (see Equation~\eqref{risk-averse utility}), a customized UCT \cite{kocsis2006bandit} in the selection phase and a specific method for predicting closing prices (see \cite{pacaud2024bidding}). It is based on two hyperparameters $\alpha$ and $N_{act}$. The first is used to arbitrate between expected utility and risk-aversion while the second corresponds to the maximum number of expanded actions per information set. This approach provide satisfying results and will be used as basis for the incomplete information case.

As an evolution of $SMS^{\alpha}$, we introduce a variant named $SMS^\alpha_{EXP3}$ in this work. This variant uses the same EXP3 algorithm as described in \cite{cowling2012information} in its selection phase, instead of UCT. More precisely, during selection, each player $i$ chooses to bid on the set of items $x_i$ with probability $\probP_i(x_i)$ at information set $I_i$:
\begin{equation}
    \probP_i(x_i)=\frac{\gamma_{I_i}}{K_{I_i}}+\frac{(1-\gamma_{I_i})}{\sum_{x'_i}\e^{\eta_{I_i}(s^\alpha_{x'_i}-s^\alpha_{x_i}) }}
\end{equation}
\noindent where $s^\alpha_{x_i}$ is an estimate of the sum of risk-averse utilities obtained after bidding on $x_i$ at $I_i$ over all search iterations where $I_i$ is encountered in the selection phase, $K_{I_i}$ is the number of possible different actions which can be selected at $I_i$ (upper-bounded by $N_{act}$), $\gamma_{I_i}$ corresponds to the probability of exploring and $\eta_{I_i}$ is a hyperparameter of the Gibbs distribution. We use the same hyperparameters suggested by Auer et al. in \cite{auer2002nonstochastic}:
\begin{equation}
    \begin{cases}
    \gamma_{I_i}=\min(1,\sqrt{\frac{K_{I_i} \ln(K_{I_i})}{(\e-1)\sum_{x'_i} n_{x'_i}}}) \\
    \eta_{I_i}=\frac{\gamma_{I_i}}{K_{I_i}}
    \end{cases}
\end{equation}
\noindent where $n_{x_i}$ is the number of times player $i$ bidded on $x_i$ at $I_i$ and $\e$ is the base of the natural logarithm.

During backpropagation, if player $i$ obtains $V^\alpha_i$ as risk-averse utility at the end of the rollout, then statistics stored for $I_i$ are updated as follows: $s^\alpha_{x_i}\gets s^\alpha_{x_i}+\frac{V^\alpha_i}{\probP_i(x_i)}$ and $n_{x_i} \gets n_{x_i}+1$.

The choice of EXP3 is motivated by the observation that the optimal policy at a simultaneous move node is often mixed \cite{cowling2012information}. Thus, using a selection strategy that returns a mixed policy, rather than a deterministic one, can significantly improve the overall performance of the algorithm.

When all search iterations have been performed, a mixed strategy for the concerned player $i$ is returned, based on the frequencies of visit counts of each expanded move. Before computing these frequencies, we remove visits caused by exploration to enhance the algorithm's performance \cite{teytaud2011upper}. We thus obtain a new visit count $n'_{x_i}$ for bidding on each set of items $x_i$. The concerned player $i$ then bids on $x_i$ with probability $\frac{n'_{x_i}}{\sum_{x'_i} n'_{x'_i}}$.

\subsection{Overview of the three determinization approaches} 

Determinization is a process that involves generating an instance of the equivalent game where all hidden information is assumed to be known to all players. Combining MCTS with determinization is a popular approach for computing good strategies in games with incomplete and imperfect information \cite{cowling2012ensemble, cowling2012information, whitehouse2011determinization}. In this section, we introduce three different determinization approaches to adapt $SMS^\alpha_{EXP3}$ to the incomplete information framework of the SAA-inc game:
\begin{itemize}
    \item \textit{Determinization through expectation} consists in using the expected value of the opponents' type distribution and apply $SMS^\alpha_{EXP3}$ to the resulting determinized SAA-inc game with complete information (\ref{determin exp}).
    \item \textit{Separate-tree determinization} consists in generating many determinized games and apply $SMS^\alpha_{EXP3}$ to each different instance. The results obtained for each instance are then combined and a final move is selected (\ref{determin sep}).
    \item \textit{Single-tree determinization} consists in generating a unique search tree and draw a different determinization at each search iteration (\ref{determin single}).
\end{itemize}

\begin{figure*}[!t]
\centering
\hspace*{-1.2cm}\subfloat[Determinization through expectation]{\includegraphics[width=3in]{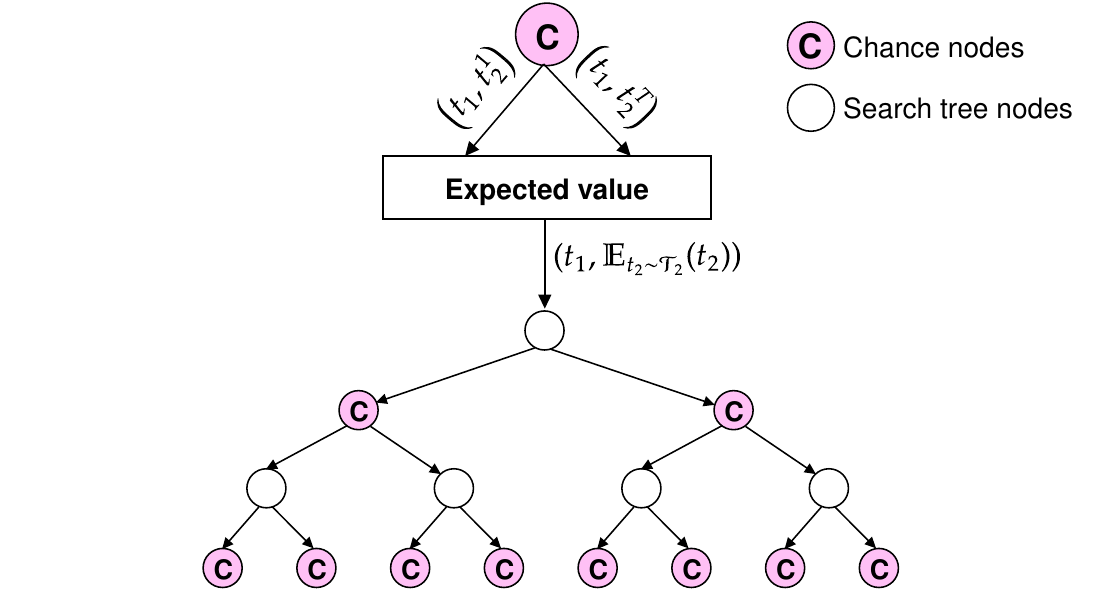}%
\label{Expect_SMS}}
~
\subfloat[Single-tree determinization]{\includegraphics[width=3in]{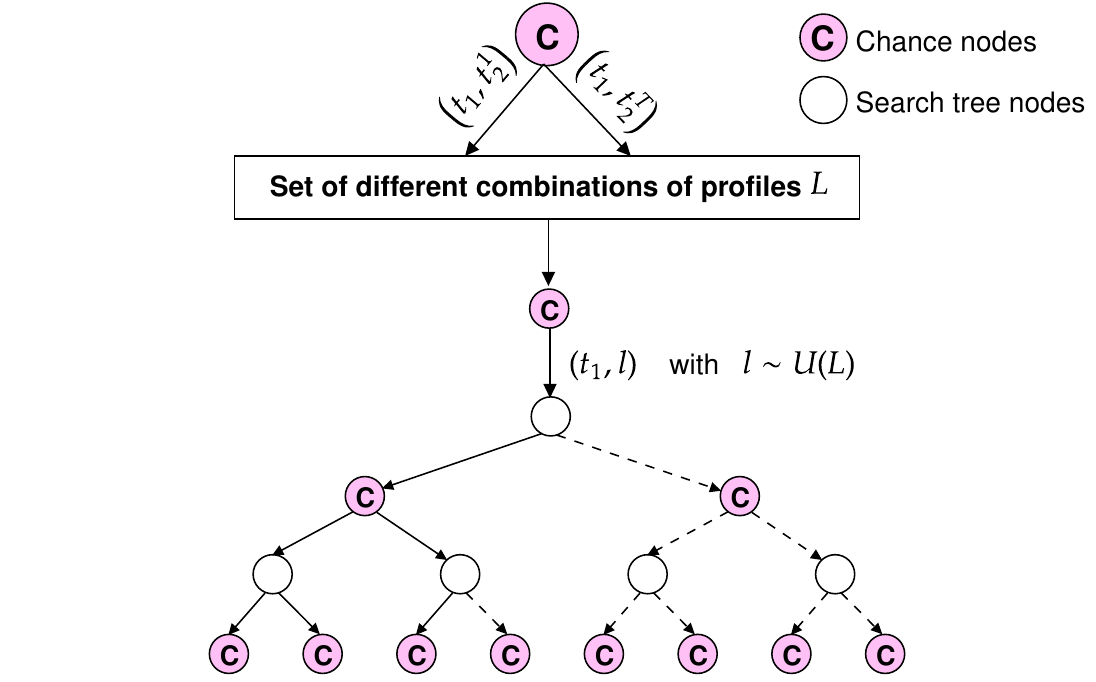}%
\label{Single_tree_deter}}

\subfloat[Separate-tree determinization]{\includegraphics[width=3in]{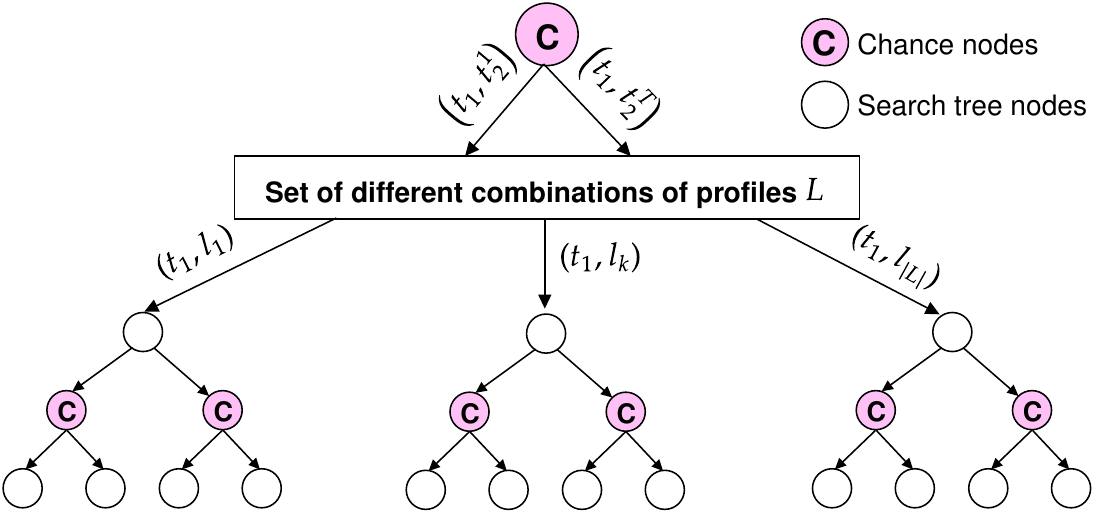}%
\label{Separate_tree_deter}}

\caption[Representation of the three different determinization approaches applied to a SAA-inc game with two players from the point of view of player $1$, with $T=|supp(\mathcal{T}_2)|$. In (b), edges with dashed lines represent moves non consistent with $l$ and are ignored.]{Representation of the three different determinization approaches applied to a SAA-inc game with two players from the point of view of player $1$, with $T=|supp(\mathcal{T}_2)|$. In (b), dashed edges represent moves non consistent with $l$ and are ignored.}
\label{Three determinization approach}
\end{figure*}

In Figure \ref{Three determinization approach}, we represent the three determinization approaches applied to a SAA-inc game with two players from the point of view of player $1$. The first chance node in each approach corresponds to the random draw by Nature of the opponent's type, i.e. player $2$. The vertices in each search tree correspond to the players' joint actions.

\subsection{Determinization through expectation}\label{determin exp}

A simple approach to apply $SMS_{EXP3}^\alpha$ to SAA-inc is to consider the corresponding bidding game with complete information, where opponents play according to the expected value of their type distribution. More formally, we suppose that an opponent $i$ with value distribution $\mathcal{F}_i$ and budget distribution $\mathcal{B}_i$ plays with value function $\overline{v_i}$ and budget $\overline{b_i}$ as follows:
\begin{equation}
    \begin{cases}
        \forall X \in \mathcal{P}(\{1,...,m\}), \overline{v_i}(X)=\E_{v\sim \mathcal{F}_i}[v(X)] \\
        \overline{b_i}=\E_{b\sim \mathcal{B}_i}[b]
    \end{cases}
\end{equation}
It is possible that $(\overline{v_i},\overline{b_i})\notin supp(\mathcal{T}i)$, meaning the considered bidding game with complete information, to which $SMS_{EXP3}^\alpha$ is applied, may not be one of the possible sub-games with complete information of the Bayes-equivalent of the initial SAA-inc game. However, for auction games, this does not pose any issue since the resulting game is still valid. This determinization approach is denoted $\overline{SMS_{EXP3}^\alpha}$. This approach is illustrated in Figure~\ref{Expect_SMS}.

\subsection{Separate-tree determinization} \label{determin sep}

The main drawback of the preceding approach is to only focus on the expectation and  ignore other aspects of the type distribution such as the variance. Thus, a simple approach to better exploit the type distribution is to generate separate trees for different determinizations and then combine the results \cite{cowling2012information}. We now show how actions are expanded at the root node and how the final move is selected.

\subsubsection{Actions expanded at the root node}

As expanding all actions at the root of each separate tree prevents in-depth inspection of promising branches, we opt to limit the number of actions expanded. While it is possible to expand different sets of actions for the concerned player at the root based on each determinization and its predicted closing prices, for fairness in combining final results, we maintain consistency by using the same set across all separate trees. We select these actions as follows. 

We denote by $L$ the set of different combinations of profiles. In each combination, the concerned player's profile is the same, i.e., every player is aware of the concerned player's true value function and budget. For each combination of profiles $l\in L$, we compute an initial prediction of closing prices $p^*(l)$ as done in \cite{pacaud2024bidding} and generate a separate tree, see Figure~\ref{Separate_tree_deter}. For the concerned player at the root, we select a maximum of $N_{act}$ actions. For the same reasons mentioned in \cite{pacaud2024bidding}, not bidding is included in the $N_{act}$ actions. The remaining $N_{act}-1$ actions correspond to the moves leading to the $N_{act}-1$ highest predicted utilities in point-price prediction bidding strategy (\textit{PP})\cite{wellman2008bidding}, extended to budget and eligibility constrained environments\cite{pacaud2022monte}, with initial prediction $\frac{1}{|L|}\sum_{l\in L} p^*(l)$. 
For all other information sets in the separate tree with combination of profiles $l\in L$, the expanded set of actions is chosen according to the highest predicted utilities in strategy \textit{PP} with initial prediction $p^*(l)$.

\subsubsection{Final move selection}

When all search iterations are performed, we obtain a mixed strategy for each combination of profiles $l$. Our final move selection is based on a variant of majority voting \cite{nijssen2012monte}. For each separate tree, an action is drawn from its final mixed strategy. The action which has been drawn the more often (and, thus, has gathered the most votes) across all separate trees is selected as the best move. If two or more moves accumulate the same highest number of votes, then the move with the highest number of visits over all trees amongst the tied moves is selected.
 
We name the resulting algorithm $DSMS^\alpha_{EXP3}$ for $Determinized \; SMS^\alpha_{EXP3}$. We provide a succinct pseudo-code in Algorithm \ref{alg:Determinized SMS_EXP3}.

\begin{algorithm}[t]
\caption{$DSMS^\alpha_{EXP3}$}\label{alg:Determinized SMS_EXP3}
\begin{algorithmic}
\State \textbf{Hyperparameters:} The risk-aversion hyperparameter $\alpha$ and the maximum number of expanded actions $N_{act}$ 
\State \textbf{Inputs computed offline:} A set of combinations of profiles $L$ and the initial prediction of closing prices $p^*(l)$ for each combination of profiles $l\in L$
\State \textit{// Selecting an expanded action set at the root node}
\State Selection of the action set $A_0$ for the concerned player at the root's information set given $N_{act}$ and $\frac{1}{|L|}\sum_{l\in L} p^*(l)$ 
\State \textit{// Initialisation} 
\State For each action $a\in A_0$, we set the total number of visits $n_a$ and the total number of votes $v_a$ to $0$
\State \textit{// Generating and running separate trees}
\For{$l\in L$}
\begin{itemize}
    \item Generate a tree with expanded set of actions $A_0$ at the root's information set for the concerned player
    \item Run $SMS^\alpha_{EXP3}$ starting from the initialised tree using determinization $l$ (the SAA-inc game considered is with complete information) 
    \item When all search iterations have been performed, withdraw the visits due to exploration and update the total number of visits $n_a$ of each action $a\in A_0$ with the number of visits $n'_{a,l}$ obtained in the tree with determinization $l$: $n_a \gets n_a+{n'}_{a,l}$ 
    \item Draw an action $a\in A_0$ with probability $\frac{n'_{a,l}}{\sum_{a'\in A_0} n'_{a',l}}$ and update its total number of votes: $v_a \gets v_a+1$ 
\end{itemize}
\EndFor 
\State \textit{// Final selection move} 
\Return Action $a\in A_0$ with the maximal number of votes $v_a$ or, in case of ties, with the maximal number of visits $n_a$ amongst the tied moves 
\end{algorithmic}
\end{algorithm}

\subsection{Single-tree determinization} \label{determin single}

In the single-tree determinization \cite{pepels2016sequential}, also named Single-Observer Information Set MCTS (SO-ISMCTS), $SMS^\alpha_{EXP3}$ runs over a single tree. At each search iteration, a combination of profiles $l$ is drawn and only the opponents' moves consistent with $l$ are considered, i.e., those compatible with their determinized budget, see Figure~\ref{Single_tree_deter}. With this method, nodes in the tree correspond to information sets from the root player's point of view and information is shared between different combinations of profiles. The search iteration phases of $SMS^\alpha_{EXP3}$ are modified as follows.

\subsubsection{Selection}

Depending on the combination of profiles $l\in L$, the available actions at a node may vary. Non-consistent moves are indeed not allowed. This issue can be seen as a multi-armed bandit problem where only a subset of the arms is available at each trial. We propose the following modifications to adapt the EXP3 algorithm to this problem, referred to as \textit{subset-armed bandit} \cite{cowling2012information}.

During the selection phase, each opponent $i$ with profile $l_i$ chooses to bid on the set of items $x_i$ with probability $\probP_i(x_i)$ at information state $I_i$:
\begin{equation}
    \probP_i(x_i)=\frac{\gamma^{l_i}_{I_i}}{K^{l_i}_{I_i}}+\frac{(1-\gamma^{l_i}_{I_i})}{\sum_{x'_i}\e^{\eta^{l_i}_{I_i}(\frac{n_{I_i}}{N_{x'_i}}s^\alpha_{x'_i}-\frac{n_{I_i}}{N_{x_i}}s^\alpha_{x_i})}}
\end{equation}
\noindent where $K^{l_i}_{I_i}$ corresponds to the number of legal moves which can be played by player $i$ at $I_i$ with profile $l_i$, $n_{I_i}$ the total number of times $I_i$ has been visited and $N_{x_i}$ the total number of times a player $i$ could have bidded on $x_i$ at $I_i$ (number of times this move was considered legal). Both hyperparameters are modified accordingly: 
\begin{equation}
    \begin{cases}
    \gamma^{l_i}_{I_i}=\min(1,\sqrt{\frac{K^{l_i}_{I_i} \ln(K^{l_i}_{I_i})}{(\e-1)n_{I_i}}}) \\
    \eta^{l_i}_{I_i}=\frac{\gamma^{l_i}_{I_i}}{K^{l_i}_{I_i}}
    \end{cases}
\end{equation}

\noindent At the end of each selection step, for all sets of items $x_i$ of $I_i$ within the budget of player $i$ according to $l_i$, we set $N_{x_i}\gets N_{x_i}+1$.

As $s^\alpha_{x_i}$ acts as an estimate of the sum of risk-averse utilities obtained after bidding on $x_i$ at $I_i$ across all search iterations where $I_i$ is encountered in the selection phase and bidding on $x_i$ is legal, $\frac{n_{I_i}}{N_{x_i}} s^\alpha_{x_i}$ provides an estimate across all search iterations where $I_i$ is encountered. This enables a fair comparison between different moves in the Gibbs distribution. Without this rescaling, moves that cannot be played with low-budget profiles would be unfairly disadvantaged when high-budget profiles are drawn. 

One slight issue with this approach is that the uniform distribution decreases in $\frac{1}{\sqrt{n_{I_i}}}$ regardless of profile $l_i$. Thus, moves which can not be played with low budget profiles have less chance of being explored than the others. Nevertheless, given the fact that the number of different profiles $l_i$ for an opponent $i$ is low, this should not be too problematic. 

During the selection phase, we apply the above modified EXP3 algorithm to the information sets of the root player's opponents. For the root player, the usual EXP3 algorithm is used as its profile remains the same across all search iterations. 

\subsubsection{Expansion phase} \label{Expansion SDSMS}

As it is the case for $SMS^\alpha_{EXP3}$, only a maximum number of actions $N_{act}$ can be expanded per information set \cite{pacaud2024bidding}. Not bidding is always included in the $N_{act}$ actions. At each search iteration, a combination of profiles $l$ is randomly drawn. For each player $i$ at information state $I_i$ with eligibility $e_i$, amongst the remaining legal unexpanded actions, we expand the action of bidding on set of items $x_i$ that leads to the highest predicted utility according to PP \cite{pacaud2024bidding} with initial prediction $p^*(l)$. Only sets of items $x_i$ which respect the eligibility and budget constraints given by $e_i$ and profile $l_i$ are considered. Statistics for each action are initialised as follows: $s^\alpha_{x_i}\gets 0$, $n_{x_i}\gets 0$ and $N_{x_i}\gets 1$.    

Unlike $SMS^\alpha_{EXP3}$ where there is a clear division between the selection phase and the expansion phase, this is not the case here. Both are intertwined. This is due to the fact that actions are added one by one in each information set and depend on the combination of profiles drawn. We can divide these in two separate phases:
\begin{itemize}
    \item \textbf{Selection of expanded nodes:} If the node is already expanded, i.e. a same configuration of temporary winners, bid prices and eligibilities has already been added to the search tree, then we check for each information state $I_i$ the two following cases: 
    \begin{itemize}
        \item If the maximum number of actions $N_{act}$ has already been reached or all legal actions have already been added to the search tree, then we apply the selection strategy described in the last subsection and draw an action from the corresponding mixed policy.
        \item Otherwise, a new action is added to the information set as described above.
    \end{itemize}
    Each player then plays its corresponding action. This process continues until a non-expanded node is reached. 
    \item \textbf{Adding a non-expanded node:} If a non-expanded node is reached, then, for each information set $I_i$, an action is expanded as detailed above and $N_{I_i}\gets 0$.
\end{itemize}

\subsubsection{Rollout}

Similarly to $SMS^\alpha_{EXP3}$, at the beginning of each rollout phase, we set $p^*_i=p^*(l)+\eta_i$ with $\eta_i\sim U([-\varepsilon,\varepsilon]^m)$ and $l$ the combination of profiles drawn at this search iteration. Each player $i$ then plays PP with initial prediction of closing prices $p^*_i$ during the entire rollout.

\subsubsection{Backpropagation}

Let $V^\alpha_i$ be the risk-averse utility obtained by player $i$ at the end of the rollout. Let $x_i$ be the set of items on which player $i$ bidded at information state $I_i$ for one of the selected nodes. The statistics stored for $I_i$ are updated as follows:

\begin{itemize}
    \item[$\ast$] $s^\alpha_{x_i}\gets s^\alpha_{x_i}+\frac{V^\alpha_i}{\probP_i(x_i)}$
    \item[$\ast$] $n_{x_i} \gets n_{x_i}+1$
    \item[$\ast$] $N_{I_i} \gets N_{I_i}+1$
\end{itemize}

In the specific case where it is the first time that $x_i$ is selected ($n_{x_i}=0$ before the backpropagation phase), we set $s^\alpha_{x_i}\gets V^\alpha_i$.

We name the resulting algorithm $SDSMS^\alpha_{EXP3}$ for \textit{Single-tree Determinization} $SMS^\alpha_{EXP3}$. We provide succinct pseudo-code in Algorithm \ref{alg:Single-tree Determinization SMS_EXP3}. 

\begin{algorithm}[t]
\caption{$SDSMS^\alpha_{EXP3}$}\label{alg:Single-tree Determinization SMS_EXP3}
\begin{algorithmic}
\State
\State \textbf{Hyperparameters:} The risk-aversion hyperparameter $\alpha$ and the maximum number of expanded actions $N_{act}$
\State \textbf{Inputs computed offline:} A set of combinations of profiles $L$ and the initial prediction of closing prices $p^*(l)$ for each combination of profiles $l\in L$ 
\For{t search iterations}
    \State \textit{// Select combination of profiles}
    \State $l\sim U(L)$ where $U$ is the uniform distribution %\\
    \State \textit{// Selection of expanded nodes}
    \State Select a path consistent with $l$ until a non-expanded node  \\ \hspace{0.45cm} is reached using the \textit{Selection of expanded} \textit{nodes} process \\ \hspace{0.45cm} with $N_{act}$ and $p^*(l)$ described in Section \ref{Expansion SDSMS}%\\
    \State \textit{// Adding a non-expanded node}
    \State Add the non-expanded node using $l$ and $p^*(l)$ %\\
    \State \textit{// Rollout}
    \State Simulate a SAA-inc game where each bidder $i$ plays a \\ \hspace{0.45cm} PP strategy with initial prediction of closing prices \\ \hspace{0.45cm} $p^*_i=p^*(l)+\eta_i$ ($\eta_i\sim U([-\varepsilon,\varepsilon]^m)$)%\\
    \State \textit{// Backpropagation}
    \State Update statistics of the selected actions using the results \\ \hspace{0.45cm} of the rollout phase
\EndFor 
\State \textit{// Final selection move} 
\State Withdraw the visits due to exploration and then compute the mixed policy for the concerned player $i$ at the root's information state $I_i$ by setting the probability $\probP_i(x_i)$ of bidding on set of items $x_i$ to $\frac{n'_{x_i}}{\sum_{x'_i} n'_{x'_i}}$. \\
\Return An action drawn from the mixed policy
\end{algorithmic}
\end{algorithm}

Note that, in case of complete information, playing $\overline{SMS_{EXP3}^\alpha}$, $DSMS^\alpha_{EXP3}$ or $SDSMS^\alpha_{EXP3}$ is equivalent to playing $SMS^\alpha_{EXP3}$.

\subsection{Tracking bid exposure} \label{bid exposure}

Past bids are an important source of information which can be used to refine one's estimate about their opponents' type distributions. We propose here an easy inference method for updating beliefs about opponents' budget distributions in a SAA-inc game. It involves tracking bid exposure \cite{bulow2009winning}, which corresponds to the sum of all bids placed by a bidder during a round, including temporary winning bids from the previous round. As bidders cannot bid above their budgets, bid exposure serves as a lower bound of their budget. By monitoring an opponent's bid exposure, one can refine beliefs about their budget distribution and narrow down their possible types. 

\begin{definition} \label{budget inference}
    \textbf{Budget inference through tracking bid exposure:} \textit{Let $\mathcal{B}_i$ be the initial budget distribution of player $i$ and $\mathcal{\hat{B}}_i$ be the inferred budget distribution of player $i$ using their bid exposure $\hat{b}_i$. Let $\probP_{\mathcal{B}_i}$ be the probability associated to $\mathcal{B}_i$ and $\probP_{\mathcal{\hat{B}}_i}$ be the probability associated to $\mathcal{\hat{B}}_i$. Thus, for all $b\in supp(\mathcal{B}_i)$, $\mathcal{\hat{B}}_i$ is defined as follows: 
    \begin{equation}
    \begin{cases}
        \text{If } b<\hat{b}_i,\; \probP_{\mathcal{\hat{B}}_i}(b)=0 \\
        \text{Otherwise,} \; \probP_{\mathcal{\hat{B}}_i}(b)=\frac{\probP_{\mathcal{B}_i}(b)}{\underset{\substack{b'\in supp(\mathcal{B}_i) \\ b'\geq \hat{b}_i}} \sum \probP_{\mathcal{B}_i}(b')}
    \end{cases}
    \end{equation}}
\end{definition}

We replace $\mathcal{B}_i$ by $\mathcal{\hat{B}}_i$ in $\overline{SMS_{EXP3}^\alpha}$, $DSMS_{EXP3}^\alpha$ and $SDSMS_{EXP3}^\alpha$ to obtain a more precise estimate of opponents' budget distribution. Profiles in $DSMS_{EXP3}^\alpha$ and $SDSMS_{EXP3}^\alpha$ are also generated using $\mathcal{\hat{B}}_i$ instead of $\mathcal{B}_i$, potentially resulting in profile changes during the auction. Ideally, one should recompute the initial prediction of closing prices $p^*(l)$ whenever a combination of profiles $l$ changes. However, due to time constraints, we maintain the same initial prediction for a combination of profiles even if an opponent's determinized budget is slightly modified. 

\section{Type Generation and Profile Selection} \label{selecting relevant combinations of profiles}
In this section, we explain our general framework to generate type distributions and select specific profiles, chosen to reduce the computational load of the proposed algorithms.

\subsection{Type Generation} \label{type generation}

We propose here a general framework for generating type distributions with a fixed level of uncertainty. For simplicity, each generation is presented with continuous distributions. %In practice, once the value distributions or budget distributions are obtained, we discretize them with a small step. 

\subsubsection{Value generation}

Many studies related to single-object auctions assume either Uniform or Log-Normal distributions for the value distribution $\mathcal{F}_i$ of player $i$ \cite{goeree2003competitive, nedelec2022learning,yuan2014empirical}. Few works have however explored value distributions for multi-item auctions, particularly in the context of SAA. One such exception is~\cite{reeves2005exploring}, who applies SAA to a scheduling problem and proposes a way to generate super-additive value functions. This setting is also adopted by Wellman et al. \cite{wellman2008bidding}. We propose here a new framework for SAA that extends our previous setting~\cite{pacaud2024bidding} to incomplete information. 

%\subsubsubsection{Complementarity distributions}

In order to build $\mathcal{F}_i$ of player $i$, we introduce a new distribution $\mathcal{G}_i^{\eta_v}$, called {\it complementarity distribution}. Consider an instance $\Gamma$ of an SAA-inc game with $n$ bidders, $m$ items, a bid increment $\varepsilon$, and a level of certainty on values of $\eta_v\in[0,1]$. Let $V$ be the maximum surplus of complementarity gained by obtaining an extra item.

\begin{definition}
    \textit{For each player $i$, the complementarity distribution $\mathcal{G}_i^{\eta_v}$ is built as follows: 
    \begin{itemize}
        \item If the set of items $X$ is a singleton, then we draw $c_X\sim U([0,\eta_v V])$ and define $\mathcal{G}_{i,X}^{\eta_v}=U([c_X,c_X+(1-\eta_v)V])$. Hence, the size of the support of $\mathcal{G}_{i,X}^{\eta_v}$ is $(1-\eta_v)V$.
        \item If the set of items $X$ is not a singleton, then we draw $c_X\sim U([0,2\eta_v V])$ and define $\mathcal{G}_{i,X}^{\eta_v}=U([c_X,c_X+2(1-\eta_v)V])$. The size of the support of $\mathcal{G}_{i,X}^{\eta_v}$ is $2(1-\eta_v)V$.
    \end{itemize}
    where $U$ is the uniform distribution.}
\end{definition}

We sample $c_X$ uniformly from $[0,\eta_v V]$ or $[0,2 \eta_v V]$, so that our framework extends our previous setting \cite{pacaud2024bidding} to incomplete information. If $\eta_v=1$ (no uncertainty), then the complementarity distributions are dirac delta distributions and the value functions are common knowledge. If $\eta_v=0$, then for every bidder $i$, $\mathcal{G}_{i,X}^{\eta_v}=U([0,V])$ if $X$ is a singleton and $\mathcal{G}_{i,X}^{\eta_v}=U([0,2V])$ otherwise.
From the complementarity distributions, we are able to draw the private value functions through the following process.

\begin{setting}\label{Value setting SAA-inc}
    \textit{Let $\mathcal{G}_i^{\eta_v}$ be the complementarity distribution of player $i$. $\mathcal{G}_i^{\eta_v}$ is common knowledge. We draw a unique vector of size $2^m$, $C_{i}\sim \mathcal{G}_i^{\eta_v}$}, and build the private valuation $v_i$ as follows: 
    \begin{equation} \label{Equation Free disposal SAA-inc}
    v_i(X)=\max_{j\in X}v_i(X\backslash \{j\})+C_{i,X}
    \end{equation}
\end{setting}

Equation \eqref{Equation Free disposal SAA-inc} ensures that all value distributions $v_i\sim \mathcal{F}_i$ respect the free disposal condition. Moreover, $C_{i,X}\sim \mathcal{G}_{i,X}^{\eta_v}$ can be interpreted as the complementarity surplus obtained when purchasing a set of items $X$ instead of purchasing the set $X\backslash \{j\}$ (with $j\in X$) which maximises $v_i$. 

\subsubsection{Budget generation}

We extend here the generation of budgets of our previous setting \cite{pacaud2024bidding} to incomplete information. 

\begin{setting}\label{budget setting SAA-inc}
    Let $\eta_b\in[0,1]$ be a level of certainty on budgets. For each player $i$, we define the size of the support of their budget distribution as $d_{\eta_b}=(1-\eta_b)(b_{max}-b_{min})$ where $b_{min}$ and $b_{max}$ are the minimal and maximal budget that a player can have. The budget distribution $\mathcal{B}_i$ is built as follows: 
    \begin{itemize}
        \item We draw $B_i\sim U([b_{min},b_{max}-d_{\eta_b}])$
        \item We define $\mathcal{B}_i=U([B_i,B_i+d_{\eta_b}])$
    \end{itemize}
    where $U$ is the uniform distribution. The private budget $b_i$ of bidder $i$ is then drawn from $\mathcal{B}_i$.
\end{setting}

If $\eta_b=1$, then the budgets are common knowledge. If in addition $\eta_v=1$, then the resulting SAA-inc bidding game is with complete information and we obtain the same generation of types as in \cite{pacaud2024bidding}. If $\eta_b=0$, then a bidder only knows that the budget of its opponents is between $b_{min}$ and $b_{max}$. 

\subsection{Profile Selection} \label{profile selection}

Due to time limitations, we are constrained to sample only few determinizations during the auction. Randomly sampling determinizations from the type distribution is not satisfactory given the small number of samples we draw. As the performance of $SDSMS^\alpha_{EXP3}$ and $DSMS^\alpha_{EXP3}$ depends heavily on the sampling quality, this could result in poor outcomes. Therefore, we opt to design opponent profiles that accurately reflect their type distribution.
Each profile is parametrised with a unique parameter $\delta$. An opponent $i$ with profile $\delta$ plays with value function $v_i^{\delta}$ and budget $b_i^{\delta}$ defined as:
\begin{equation}
    \begin{cases}
        \forall X \in \mathcal{P}(\{1,...,m\}), v_i^{\delta}(X)=\E_{v\sim \mathcal{F}_i}(v(X))+\\ \hspace{4.5cm}\delta\sqrt{\mathbb{V}_{v\sim \mathcal{F}_i}(v(X))}\\
        b_i^{\delta}=\E_{b\sim \mathcal{B}_i}(b)+\delta\sqrt{\mathbb{V}_{b\sim \mathcal{B}_i}(b)}
    \end{cases}
\end{equation}
It is plausible that a profile $(v_i^{\delta},b_i^\delta)\notin supp(\mathcal{T}_i)$. Nevertheless, we notice that by selecting well chosen profiles, our method provides a better representation of the type distributions than selecting few random samples. 

The aggressiveness of the opponents in the determinized game is influenced by $\delta$. Higher values of $\delta$ lead to more aggressive play, while lower values result in more risk-averse behaviour. When sampling high values and budgets, the algorithms become more risk-averse, aiming to mitigate the risk of exposure. Sampling lower values and budgets prompts algorithms to bid more aggressively, taking advantage of opponents' weaker positions. To ensure a balanced representation of opponent types, symmetric values $\delta$ and $-\delta$ are chosen, resulting in a balanced mix of profiles for opponents.

\begin{example}
    Consider an SAA-inc game between $4$ players. A possible combination of profiles is to assign to each opponent a profile $(v_i^{\delta},b_i^\delta)$ with $\delta \in \{-1,0,1\}$. Thus, each opponent can have one of the three profiles. Therefore, the set of combinations of profiles $L$ has $27$ elements. 
\end{example}

\section{Numerical Results} \label{extensive experiments}

In this section, we evaluate the performance of our three determinization approaches by comparing them to state-of-the-art algorithms in SAA-inc games across various levels of certainty. 

\subsection{Benchmark}

We compare our three determinization approaches to the four following state-of-the-art algorithms: 

\begin{itemize}
    \item EDPE: A \textit{PP} strategy using expected-demand price equilibrium \cite{wellman2008bidding} as initial prediction of closing prices.
    \item EPE: A \textit{PP} strategy using expected price equilibrium \cite{wellman2008bidding} as initial prediction of closing prices.
    \item SB: Straightforward bidding \cite{milgrom2000putting} which is equivalent to a \textit{PP} strategy using the null vector of prices as initial prediction.
    \item SCPD: A distribution price prediction strategy using self-confirming price distribution \cite{wellman2008bidding} as initial distribution prediction.
    \item $CSMS^\alpha_{EXP3}$ (or Cheating $SMS^\alpha_{EXP3}$): An oracle strategy, not subject to uncertainty, that perfectly observes the game state and applies $SMS^\alpha_{EXP3}$.
\end{itemize}

\noindent The first four strategies are known as perceived-price bidding strategies (\textit{PPB}). We set $S_{PPB}=$\{$EDPE$, $EPE$, $SB$, $SCPD$\} and $S_{SMS}=$\{$\overline{SMS_{EXP3}^\alpha}$, $DSMS_{EXP3}^\alpha$, $SDSMS_{EXP3}^\alpha$\}. Note that all strategies in $S_{SMS}$ are equivalent to $CSMS_{EXP3}^\alpha$ in the complete information framework.

\subsection{Experimental setting}

We study SAA-inc instances with $n=3$, $m=9$ and $\varepsilon=1$. Each experimental result has been run on 1000 different SAA-inc instances. Value and budget distributions are generated using Settings \ref{Value setting SAA-inc} and \ref{budget setting SAA-inc} with $V=5$, $b_{min}=10$, $b_{max}=40$, $\eta_v\in \{0,0.5,0.8\}$ and $\eta_b=\eta_v$. 
All experiments are run on a server with Intel\textregistered Xeon\textregistered E5-2699 v4 2.2GHz processors. $DSMS_{EXP3}^\alpha$ and $SDSMS_{EXP3}^\alpha$ rely on a set of combinations of profiles $L$. We assign each opponent $i$ a profile $(v_i^\delta,b_i^\delta)$ with $\delta\in \{-1,0,1\}$. We show in Table~\ref{Hyperparameters SAA-inc} the hyperparameters $\alpha$ used by each algorithm for each level of certainty. The maximum number of expanded actions per information set is $N_{act}=20$, except for $DSMS_{EXP3}^\alpha$, for which $N_{act}=10$. $DSMS_{EXP3}^\alpha$ generates indeed $9$ separate trees instead of one for the other approaches. Given the time constraint, it must thus create smaller search trees. Both hyperparameters are selected through grid-search as in~\cite{pacaud2024bidding}. 
\begin{table}[h]
    \centering
    \caption{Hyperparameter $\alpha$ used for each approach based on $SMS_{EXP3}^\alpha$ and each level of certainty $\eta_v=\eta_b$}
    \resizebox{.3\textwidth}{!}{
    \begin{tabular}{|c c c c|} 
    \hline
    $\eta_v$ & $0$ & $0.5$ & $0.8$\\ [0.5ex] 
    \hline
    $CSMS_{EXP3}^\alpha$ & $0.8$ & $0.8$ &  $0.8$\\
    $\overline{SMS_{EXP3}^\alpha}$ & $0.2$ & $0.8$ & $0.8$ \\ 
    $DSMS_{EXP3}^\alpha$ & $0.1$ & $0.2$ &  $0.2$\\
    $SDSMS_{EXP3}^\alpha$ & $0.7$ & $0.5$ &  $0$\\
    \hline
    \end{tabular}
    }
    \vspace{0.1 cm}
    \label{Hyperparameters SAA-inc}
\end{table}
Each algorithm is given respectively $100$ seconds of thinking time except $DSMS_{EXP3}^\alpha$ where we give $50$ seconds per tree. Initial prediction of closing prices for each algorithm are done offline. 
To facilitate our analysis, measures on all performance indicators except the expected utility are obtained by confronting a strategy $A$ to a strategy $B$ and averaging the results over the two possible strategy profiles: ($A$,$B$,$B$) and ($A$,$A$,$B$). 

\subsection{Expected utility}

To simplify our study, we use the same empirical game analysis approach than Wellman et al. in \cite{wellman2008bidding}. We consider the normal form game in expected utility where each player has the choice between  a strategy $A$ and a strategy $B$. 

\begin{figure*}[t]
\centering
\subfloat[$SDSMS_{EXP3}^\alpha$ vs SCPD]{\includegraphics[width=2in]{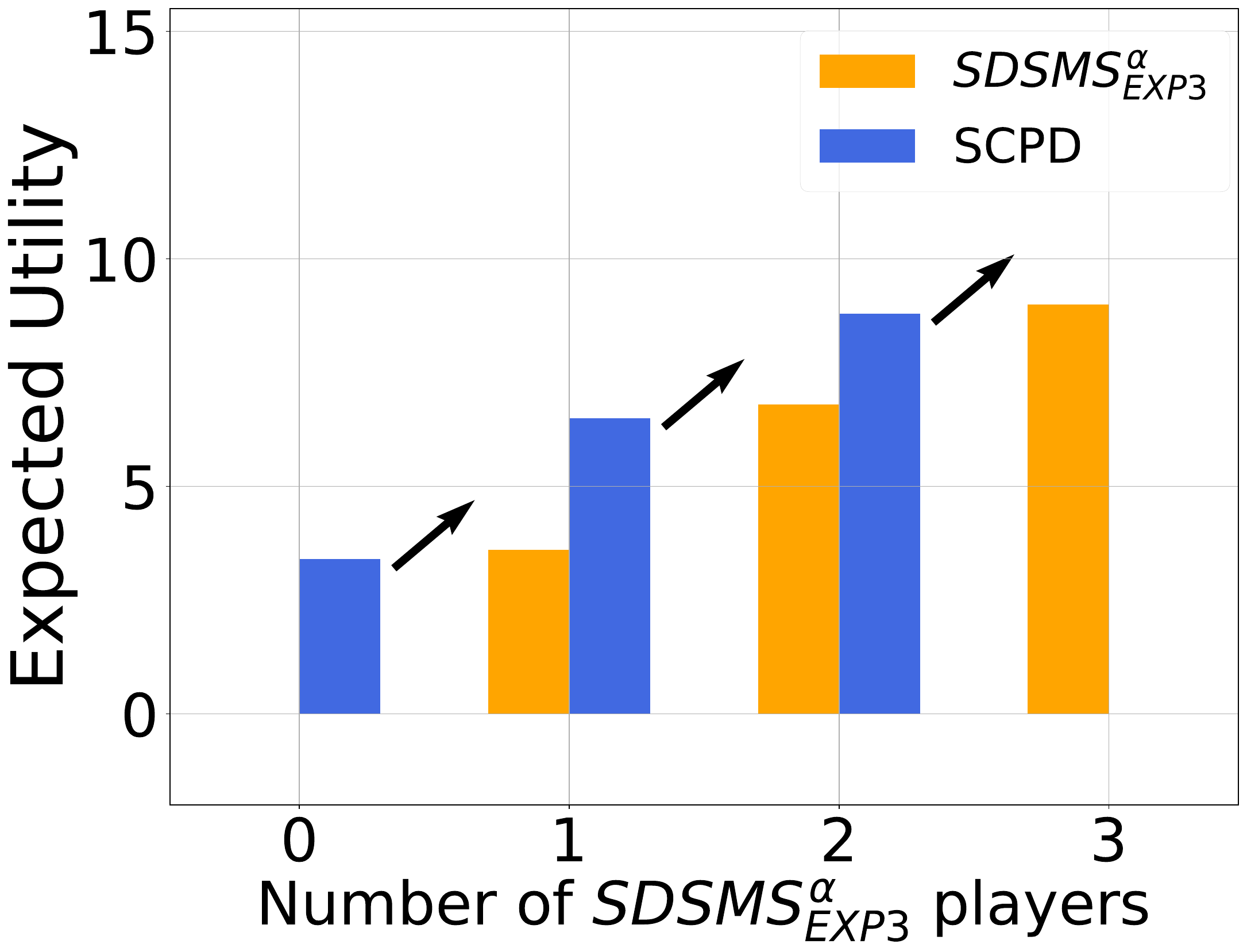}%
\label{SDSMS vs SCPD}}
~
\subfloat[$SDSMS_{EXP3}^\alpha$ vs SB]{\includegraphics[width=2in]{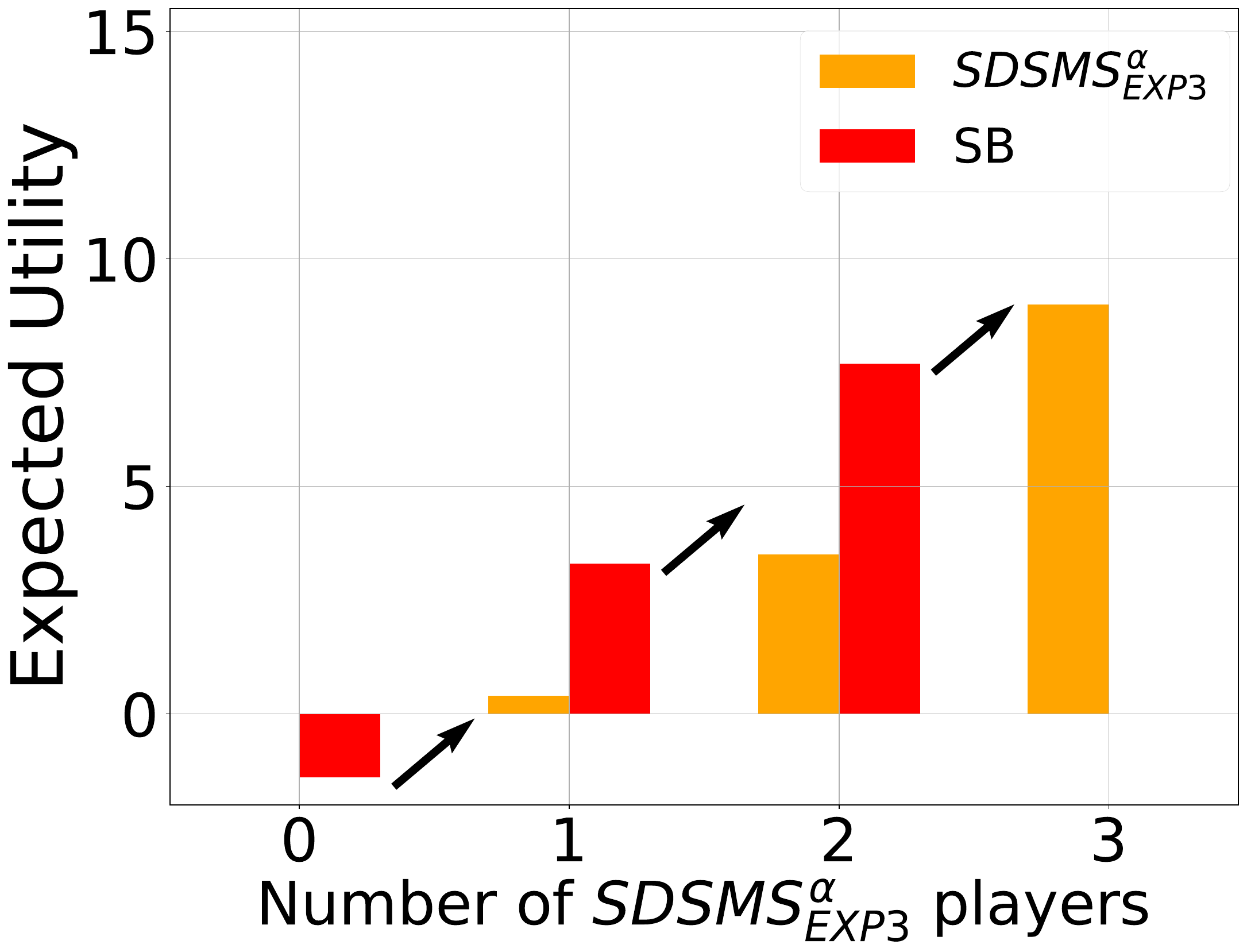}%
}
~
\subfloat[$SDSMS_{EXP3}^\alpha$ vs EPE]{\includegraphics[width=2in]{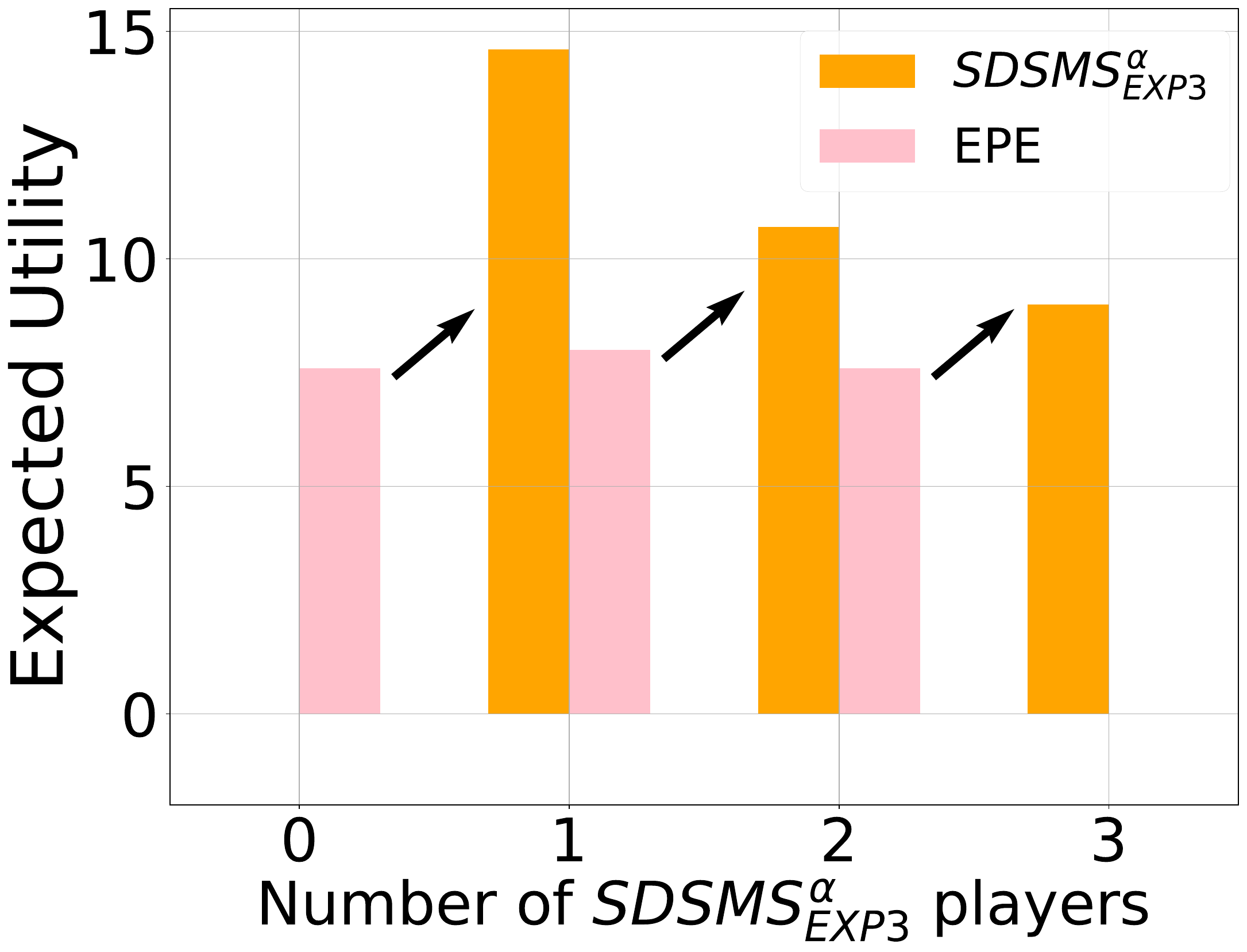}% 
\label{SDSMS vs EPE}}

\subfloat[$DSMS_{EXP3}^\alpha$ vs SCPD]{\includegraphics[width=2in]{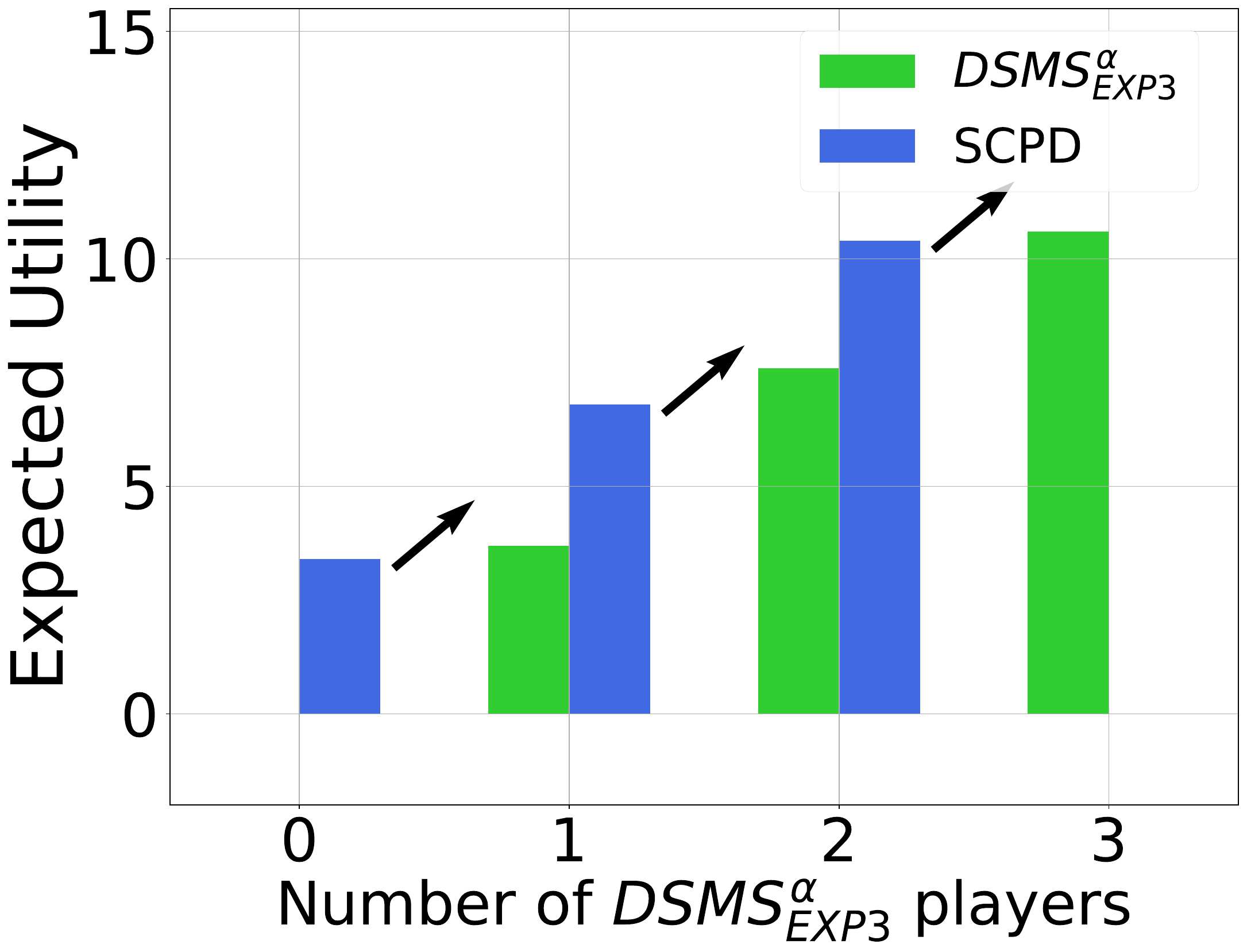}%
\label{DSMS vs SCPD}}
~
\subfloat[$DSMS_{EXP3}^\alpha$ vs SB]{\includegraphics[width=2in]{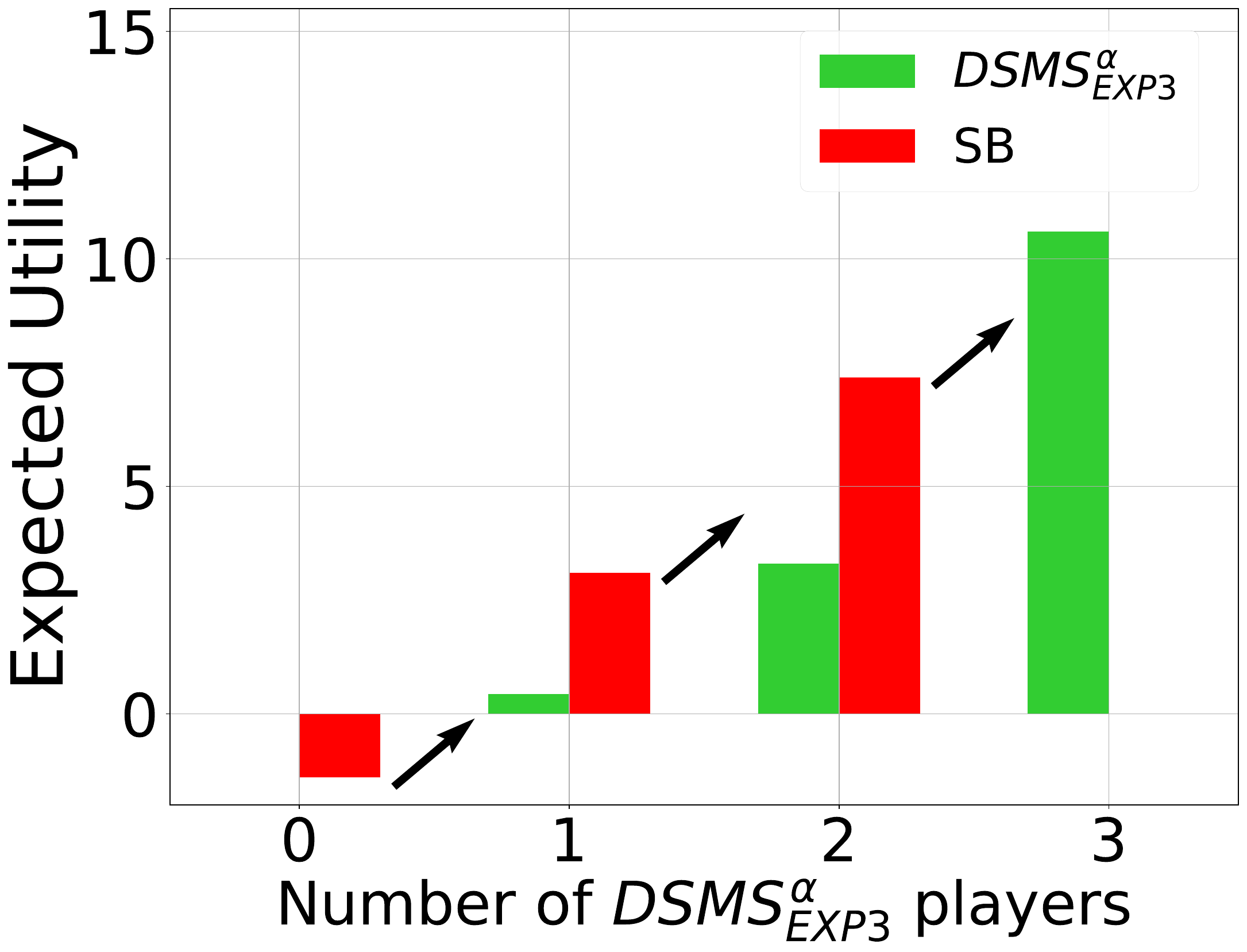}%
}
~
\subfloat[$DSMS_{EXP3}^\alpha$ vs EPE]{\includegraphics[width=2in]{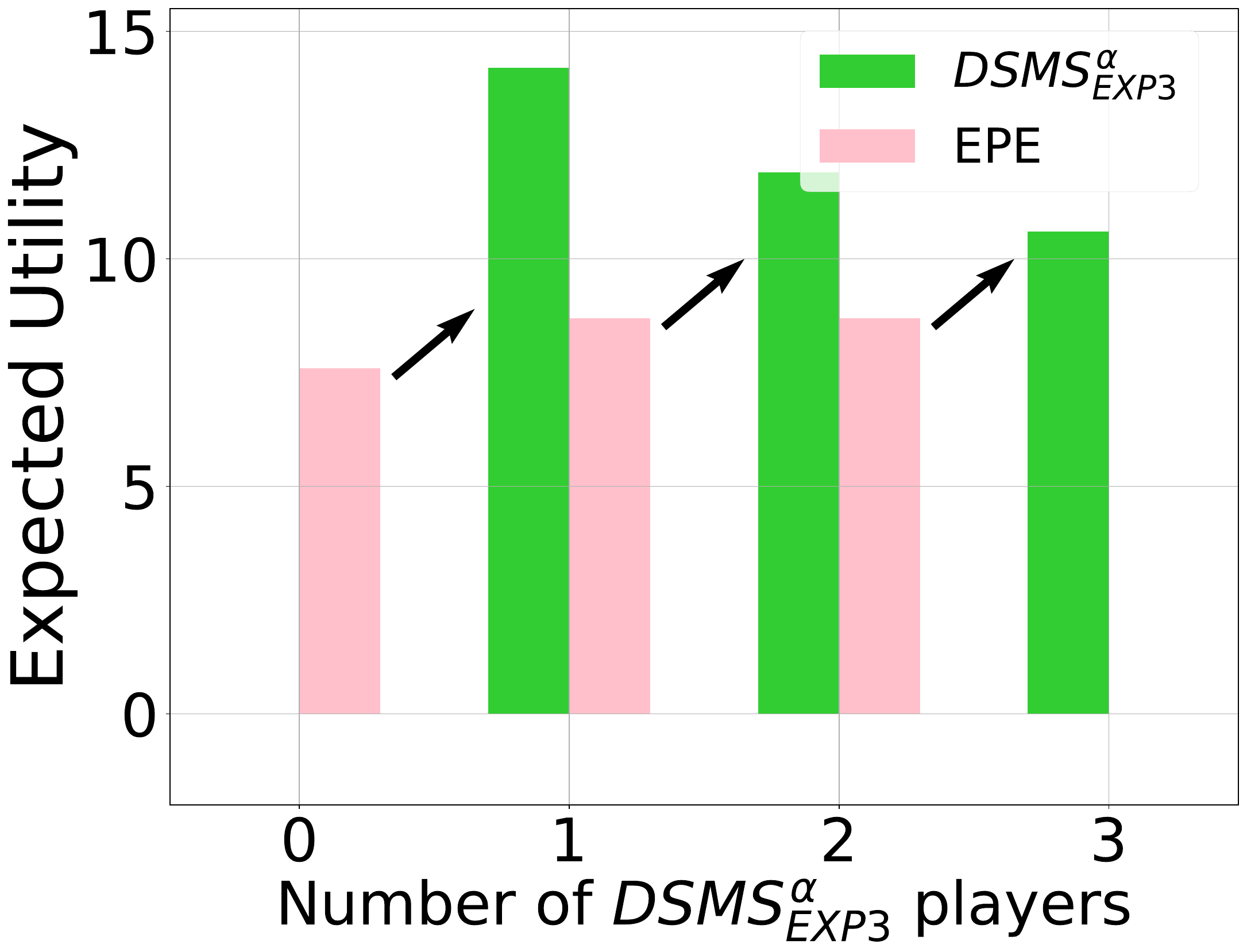}%
}

\subfloat[$\overline{SMS_{EXP3}^\alpha}$ vs SCPD]{\includegraphics[width=2in]{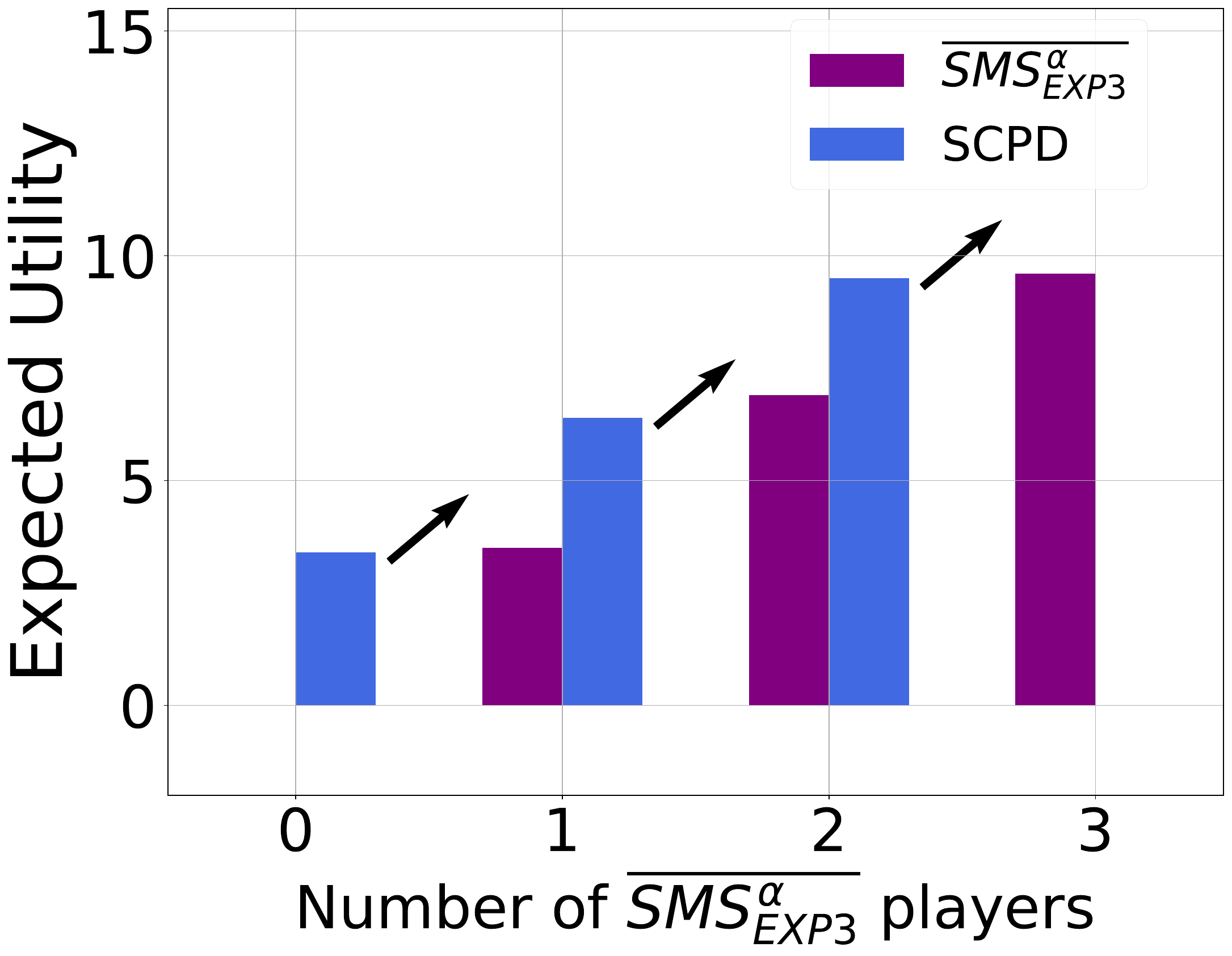}%
\label{Expect_SMS vs SCPD}}
~
\subfloat[$\overline{SMS_{EXP3}^\alpha}$ vs SB]{\includegraphics[width=2in]{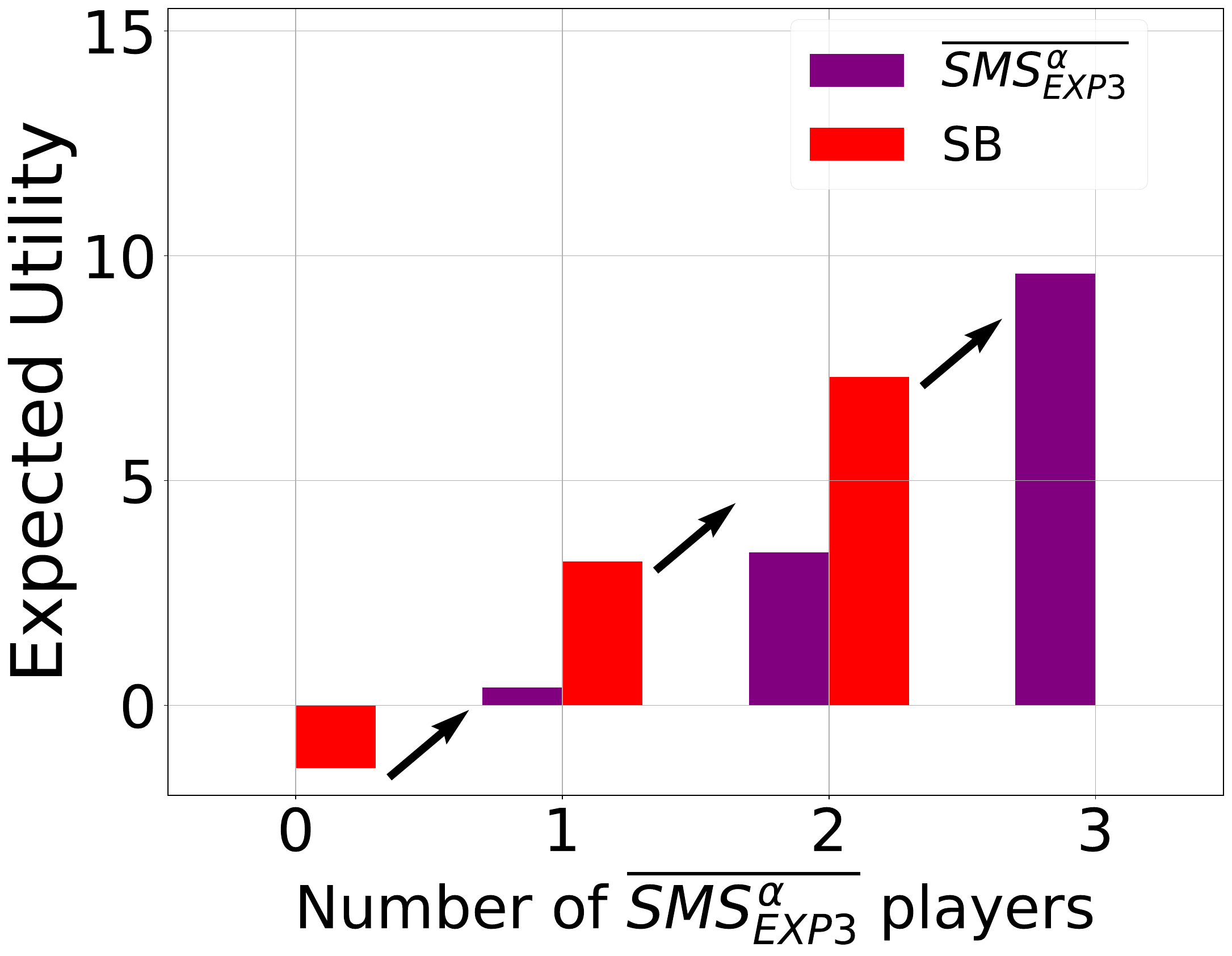}%
}
~
\subfloat[$\overline{SMS_{EXP3}^\alpha}$ vs EPE]{\includegraphics[width=2in]{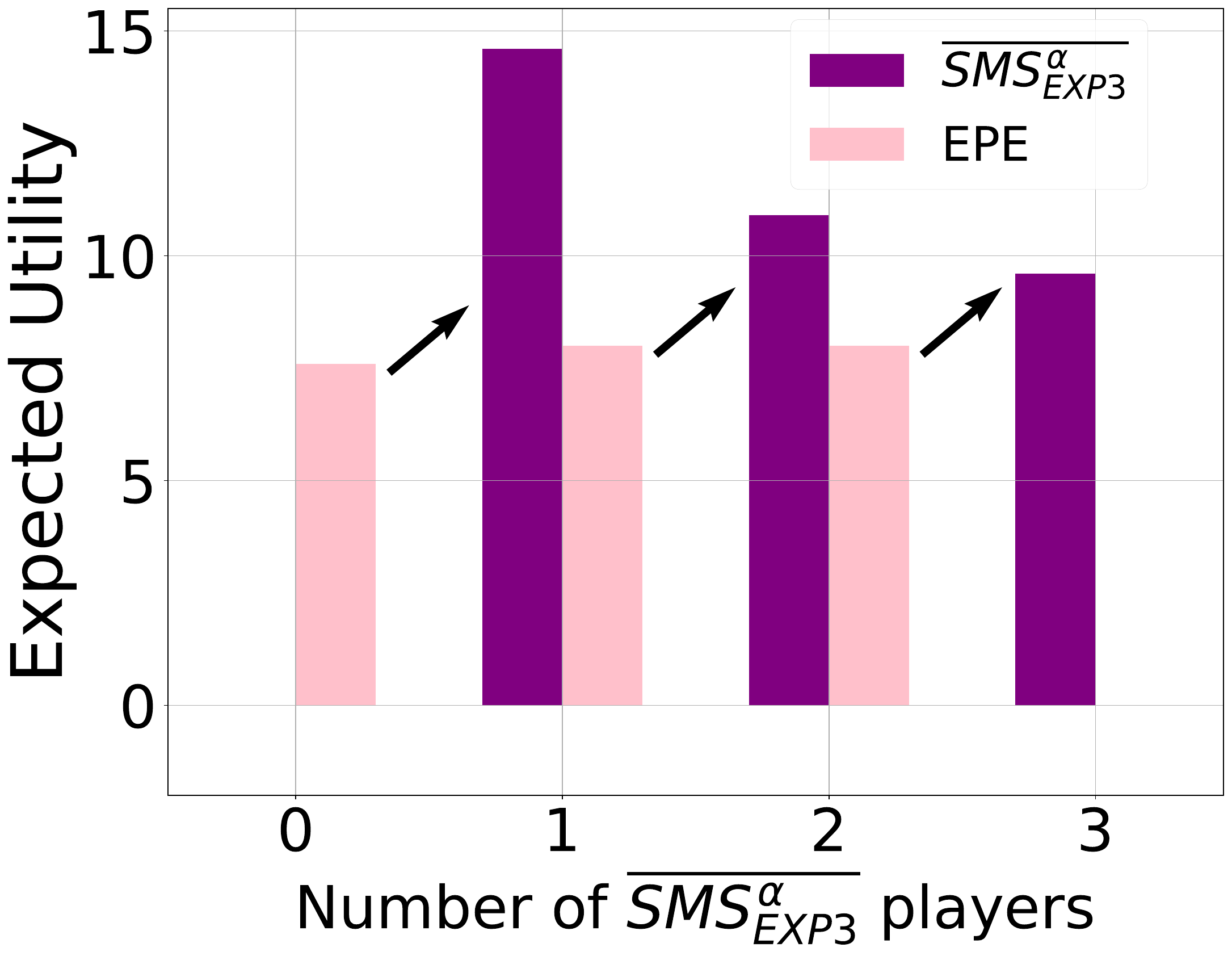}%
}
\caption{Comparing our determinization approaches with PPB approaches through normal-form SAA-inc games in expected utility with a level of certainty of $(\eta_v,\eta_b)=(0.5,0.5)$.}
\label{Expected utility SAA-inc}
\end{figure*}

\subsubsection{Comparing our determinization approaches with PPB approaches}

In Figure \ref{Expected utility SAA-inc}, we represent all empirical games for a level of certainty of $(\eta_v, \eta_b)=(0.5,0.5)$ where players have the choice between a strategy $A\in S_{SMS}$ and a strategy $B\in S_{PPB}$ (except EDPE which is strictly dominated by all strategies). Thus, we observe that for each empirical game, deviating from $B\in S_{PPB}$ to $A\in S_{SMS}$ is always profitable, in term of expected utility. Hence, for any strategy $A\in S_{SMS}$, the strategy profile ($A$, $A$, $A$) is a Nash equilibrium of the normal form SAA-inc game in expected utility with set of strategies \{$A$, SB, EPE, EDPE, SCPD\}. The same conclusion can be drawn for the two other levels of certainty proposed in the settings. Furthermore, when all bidders play the same strategy, strategies in $S_{SMS}$ consistently yield higher expected utilities than those in $S_{PPB}$. 

\subsubsection{Comparing our three determinization approaches}

\begin{figure*}[t]
\centering
\subfloat[$SDSMS_{EXP3}^\alpha$ vs $\overline{SMS_{EXP3}^\alpha}$]{\includegraphics[width=2.1in]{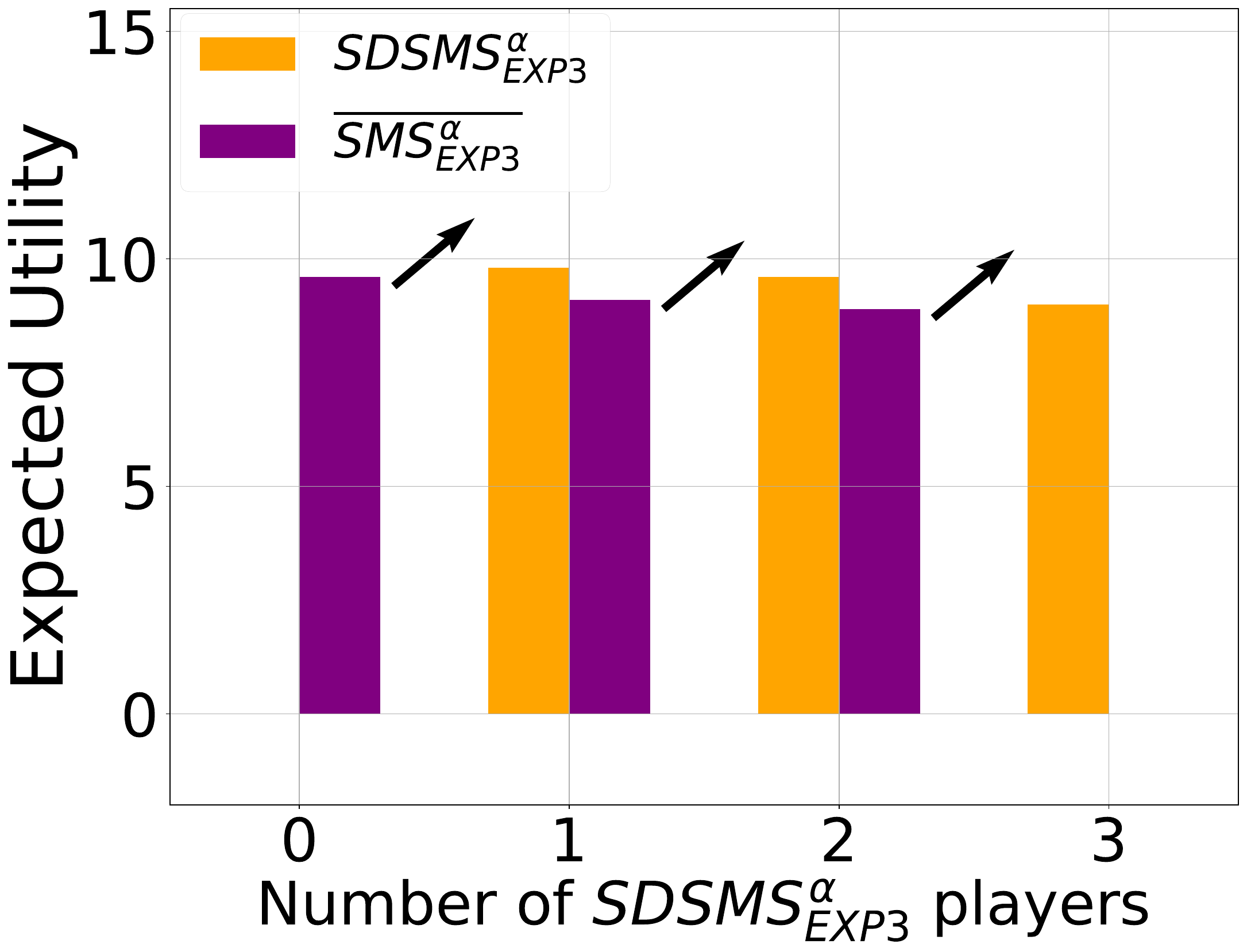}%
}
~
\subfloat[$DSMS_{EXP3}^\alpha$ vs $\overline{SMS_{EXP3}^\alpha}$]{\includegraphics[width=2.1in]{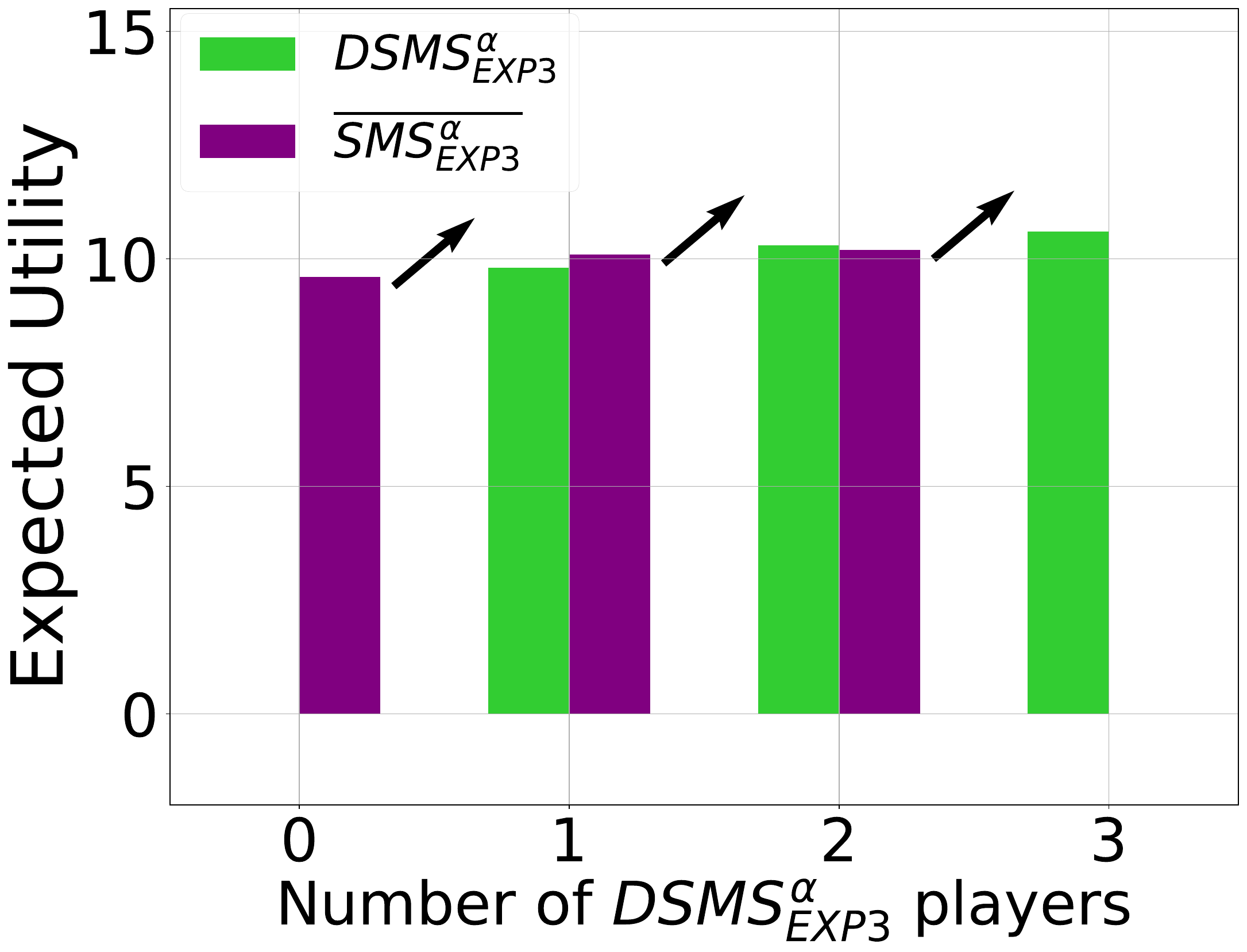}%
}
~
\subfloat[$SDSMS_{EXP3}^\alpha$ vs $DSMS_{EXP3}^\alpha$]{\includegraphics[width=2.1in]{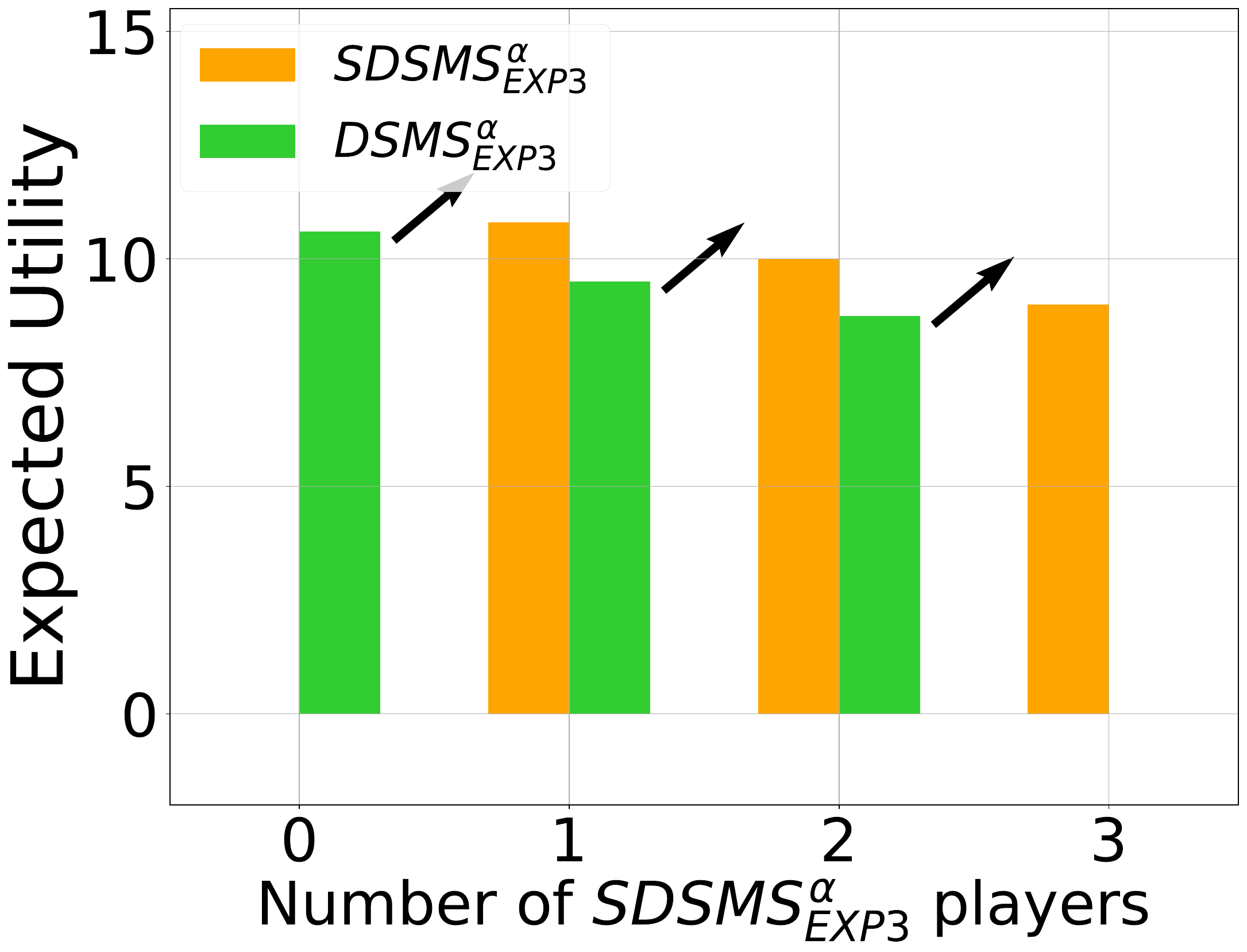}%
\label{SDSMS vs DSMS}}
\caption{Comparing our three determinization approaches through normal-form SAA-inc games in expected utility with a level of certainty of $(\eta_v,\eta_b)=(0.5,0.5)$.} 
\label{Expected utility MCTS SAA-inc}
\end{figure*}

In Figure \ref{Expected utility MCTS SAA-inc}, we represent all empirical games between our three determinization approaches for $(\eta_v,\eta_b)=(0.5,0.5)$. Deviating from $\overline{SMS_{EXP3}^\alpha}$ or $DSMS_{EXP3}^\alpha$ to $SDSMS_{EXP3}^\alpha$ is always profitable. This is also the case for the two other levels of certainty. Thus, the profile of strategy ($SDSMS_{EXP3}^\alpha$, $SDSMS_{EXP3}^\alpha$, $SDSMS_{EXP3}^\alpha$) is a Nash equilibrium of the normal-form SAA-inc game in expected utility with strategy set $S_{SMS}$. Moreover, in each empirical game, $\overline{SMS_{EXP3}^\alpha}$ is strictly dominated by the two other determinization approaches. However, the deviations never generate more than a $6\%$ increase in expected utility, regardless of the level of certainty. This unexpected small difference could be attributed to two factors. First, the value distribution for obtaining a set of items appears to converge to a symmetrical unimodal distribution as the set size increases. This tendency may favor methods like $\overline{SMS_{EXP3}^\alpha}$, which rely solely on the expected value for decision-making. Secondly, the limited number of combination of profiles used in $SDSMS_{EXP3}^\alpha$ and $DSMS_{EXP3}^\alpha$ may also contribute to their inability to significantly outperform $\overline{SMS_{EXP3}^\alpha}$. When all bidders play the same strategy, $DSMS_{EXP3}^\alpha$ always achieves a higher expected utility than $SDSMS_{EXP3}^\alpha$ or $\overline{SMS_{EXP3}^\alpha}$ for all levels of certainty. 

\subsubsection{Impact of uncertainty on our three determinization approaches}

\begin{figure*}[t]
\centering
\subfloat[$CSMS_{EXP3}^\alpha$ vs $SDSMS_{EXP3}^\alpha$]{\includegraphics[width=2.1in]{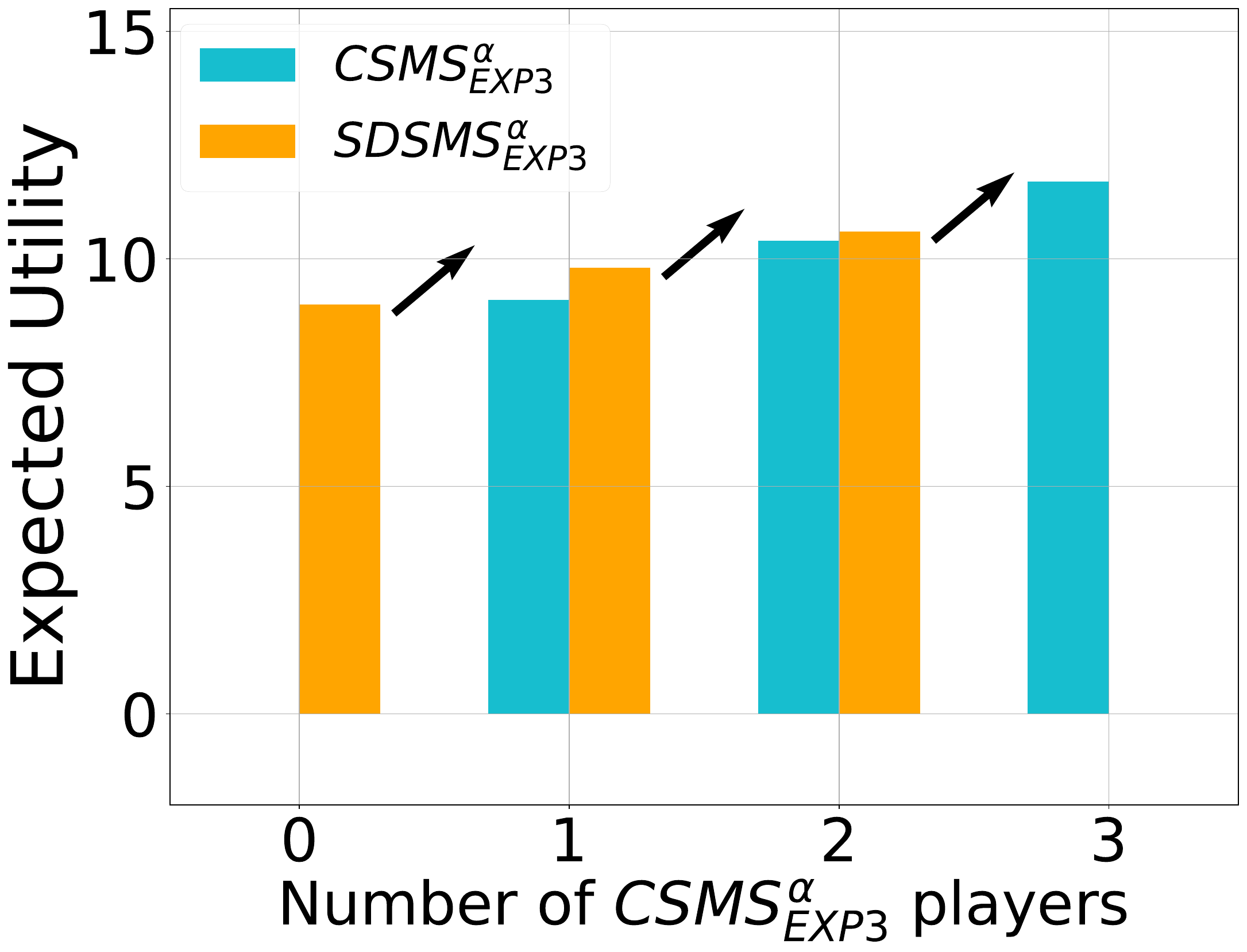}%
}
~
\subfloat[$CSMS_{EXP3}^\alpha$ vs $DSMS_{EXP3}^\alpha$]{\includegraphics[width=2.1in]{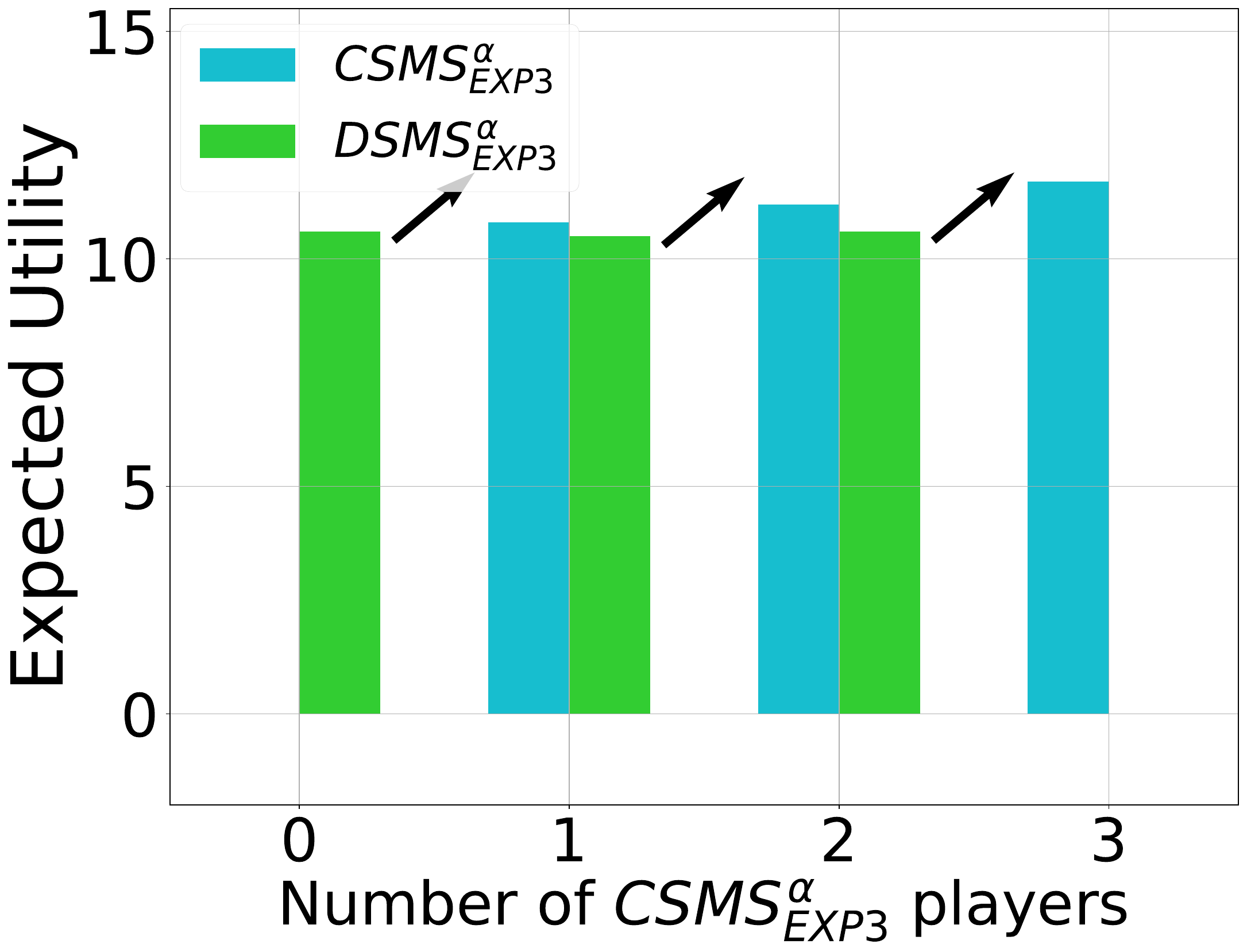}%
}
~
\subfloat[$CSMS_{EXP3}^\alpha$ vs $\overline{SMS_{EXP3}^\alpha}$]{\includegraphics[width=2.1in]{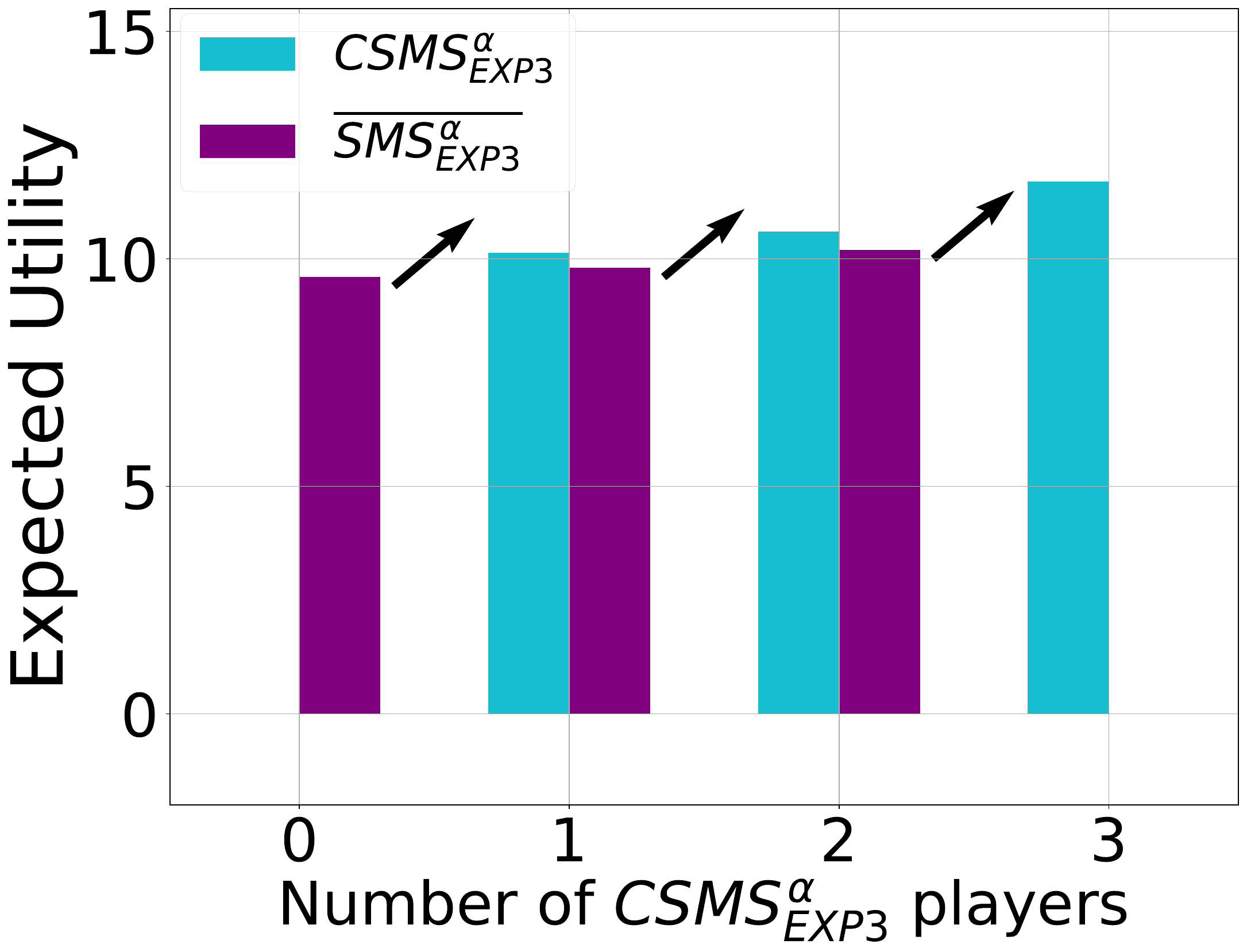}%
}
\caption{Comparing our three determinization approaches to $CSMS_{EXP3}^\alpha$ through normal-form SAA-inc games in expected utility with a level of certainty of $(\eta_v,\eta_b)=(0.5,0.5)$.} 
\label{Expected utility CSMS SAA-inc}
\end{figure*}

In Figure \ref{Expected utility CSMS SAA-inc}, we represent each empirical game between $A\in S_{SMS}$ and $CSMS_{EXP3}^\alpha$ for $(\eta_v,\eta_b)=(0.5,0.5)$. As anticipated, deviating to $CSMS_{EXP3}^\alpha$ in each empirical game is always profitable. This highlights that having accurate estimates of opponents' types increases one's performance. Uncertainty logically impacts the level of expected utility. This can mainly be explained by two factors. First, uncertainty impacts the quality of sampling of our determinization approaches as the risk-averse utilities backpropagated and legal moves considered for one's opponents are plausibly incorrect. Secondly, uncertainty also impacts our initial prediction of closing prices. Whereas $CSMS_{EXP3}^\alpha$ bidders share the same initial prediction of closing prices, these significantly vary between bidders playing the same strategy $A\in S_{SMS}$, especially when the level of certainty is low. 

\subsection{Exposure}

\subsubsection{Comparing our determinization approaches with PPB approaches}

\begin{figure*}[h!]
\centering
\subfloat[Expected exposure]{\includegraphics[width=2.3in]{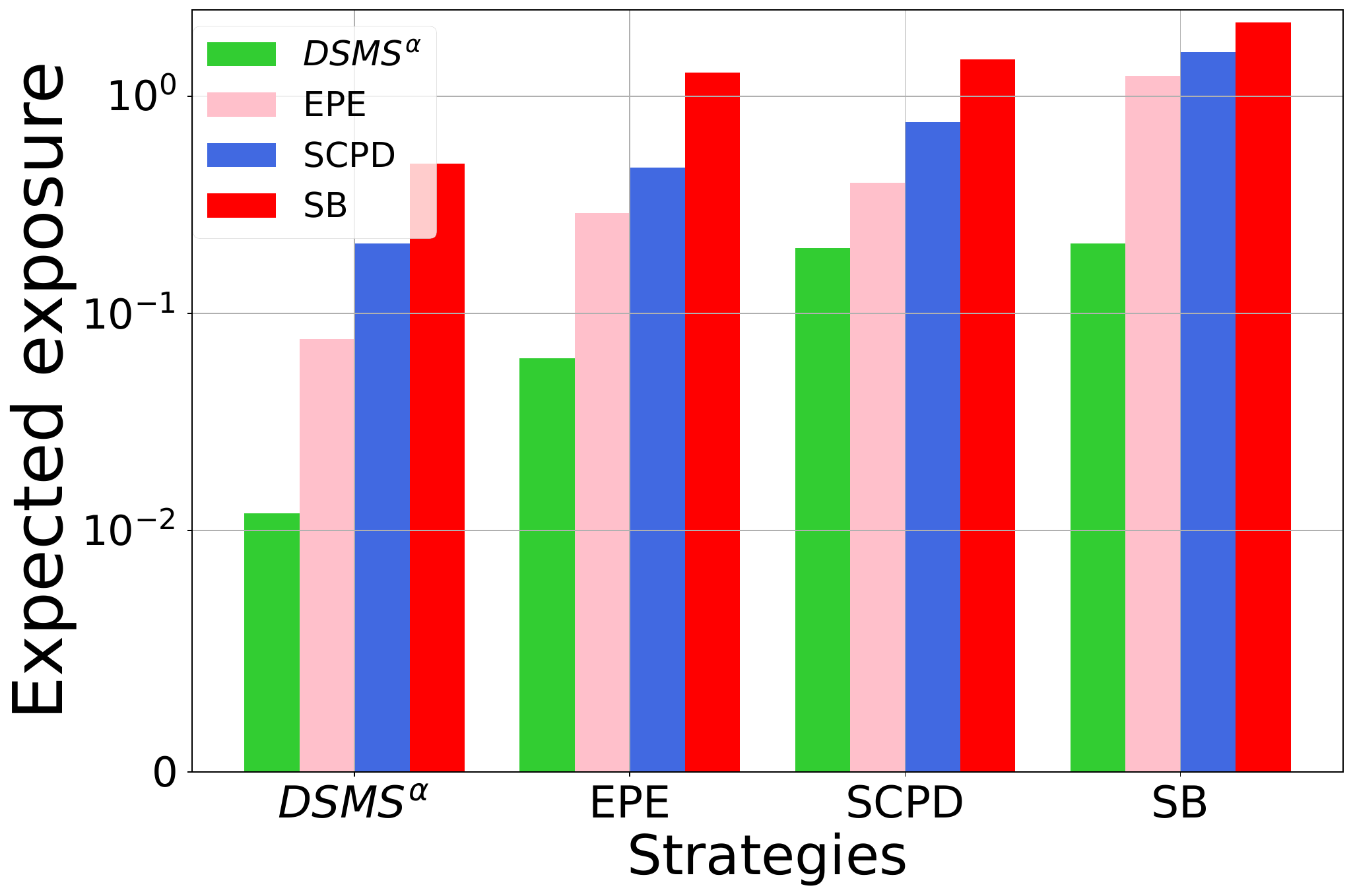}%
\label{expected exposure PP}}
\hfil
\subfloat[Exposure frequency]{\includegraphics[width=2.3in]{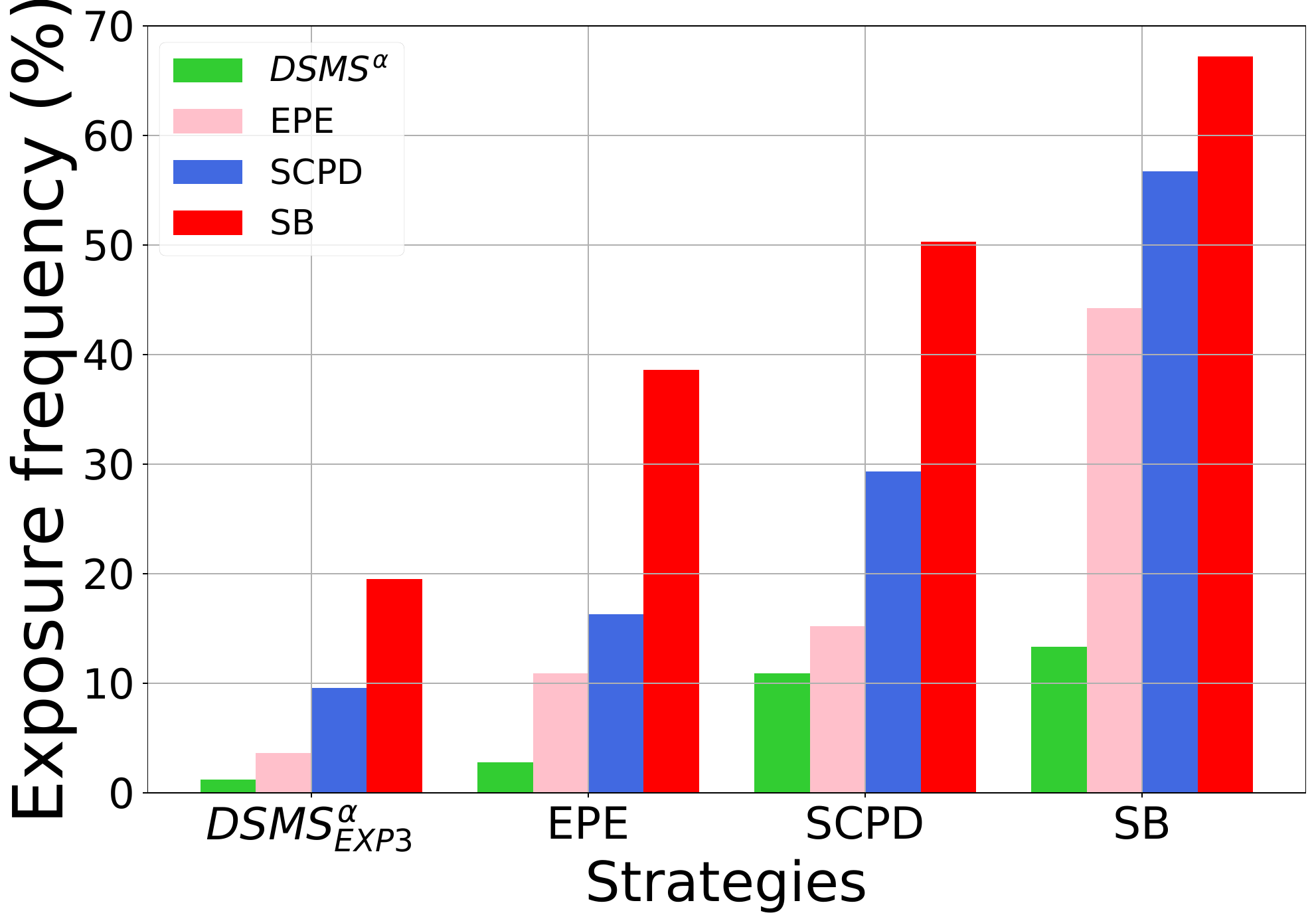}%
\label{exp_freq PP}}
\caption{Comparing the exposure suffered by $DSMS_{EXP3}^\alpha$ with the exposure suffered by PPB approaches with a level of certainty $(\eta_v,\eta_b)=(0.5,0.5)$ through two performance indicators: the expected exposure and the exposure frequency}
\label{fig:Exposure PP}
\end{figure*}

Figure~\ref{fig:Exposure PP} shows the expected exposure and the exposure frequency experienced by each strategy $A\in S_{PPB}\cup \{DSMS_{EXP3}^\alpha\}$ against every strategy $B$ on the x-axis when $(\eta_v,\eta_b)=(0.5,0.5)$. For example, when $A=DSMS_{EXP3}^\alpha$ and $B=\text{SB}$, the expected exposure by $DSMS_{EXP3}^\alpha$ against SB is $0.21$, corresponding to the green bar above index SB on the x-axis. We show here only the results related to $DSMS_{EXP3}^\alpha$ because it achieves the highest expected exposure among our determinization approaches. Nevertheless, $DSMS_{EXP3}^\alpha$ has a considerably lower expected exposure and exposure frequency than PPB strategies. When all bidders adopt the same strategy, the profile corresponding to $DSMS_{EXP3}^\alpha$ yields also the lowest expected exposure and exposure frequency. Note also that the expected exposure and the exposure frequency of the strategies in $S_{SMS}$ against each other do not present notable performance differences (not shown here). 

\subsubsection{Impact of uncertainty on our three determinization approaches}

\begin{figure}[t]
\centering
\includegraphics[width=0.7\linewidth]{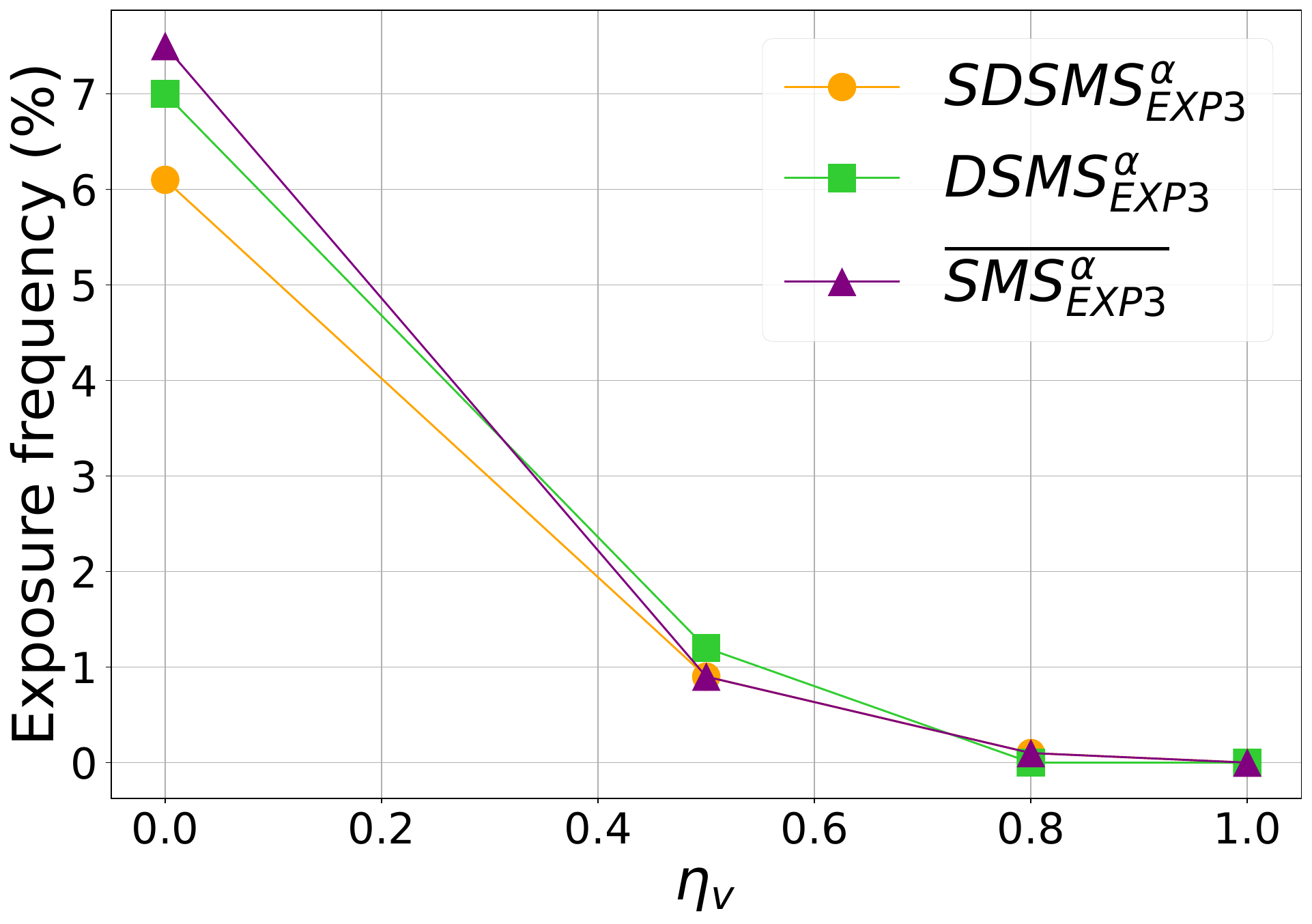}
\caption{Exposure frequency ($\%$) of our three determinization approaches when all players play the same strategy with different levels of certainty ($\eta_v=\eta_b$).} 
\label{fig:Exposure freq uncertainty SAA-inc}
\end{figure}

In Figure \ref{fig:Exposure freq uncertainty SAA-inc}, we illustrate the exposure frequency of our three determinization approaches when all bidders employ the same strategy across different levels of certainty. We observe that uncertainty affects our determinization approaches similarly, with exposure frequency sharply decreasing from $\eta_v=0$ to $\eta_v=0.5$ and then gradually approaching $0$ as $\eta_v$ approaches $1$. This suggests that coordination among bidders playing the same strategy becomes riskier with increasing uncertainty. To summarize, our determinization approaches outperforms the state-of-the-art in terms of exposure. Their performance decreases with an increasing level uncertainty. We didn't observe significant differences between the proposed solutions with this criterion.

\subsection{Own price effect}

\begin{figure*}[t]
\centering
\subfloat[Average price paid per item won]{\includegraphics[width=2.3in]{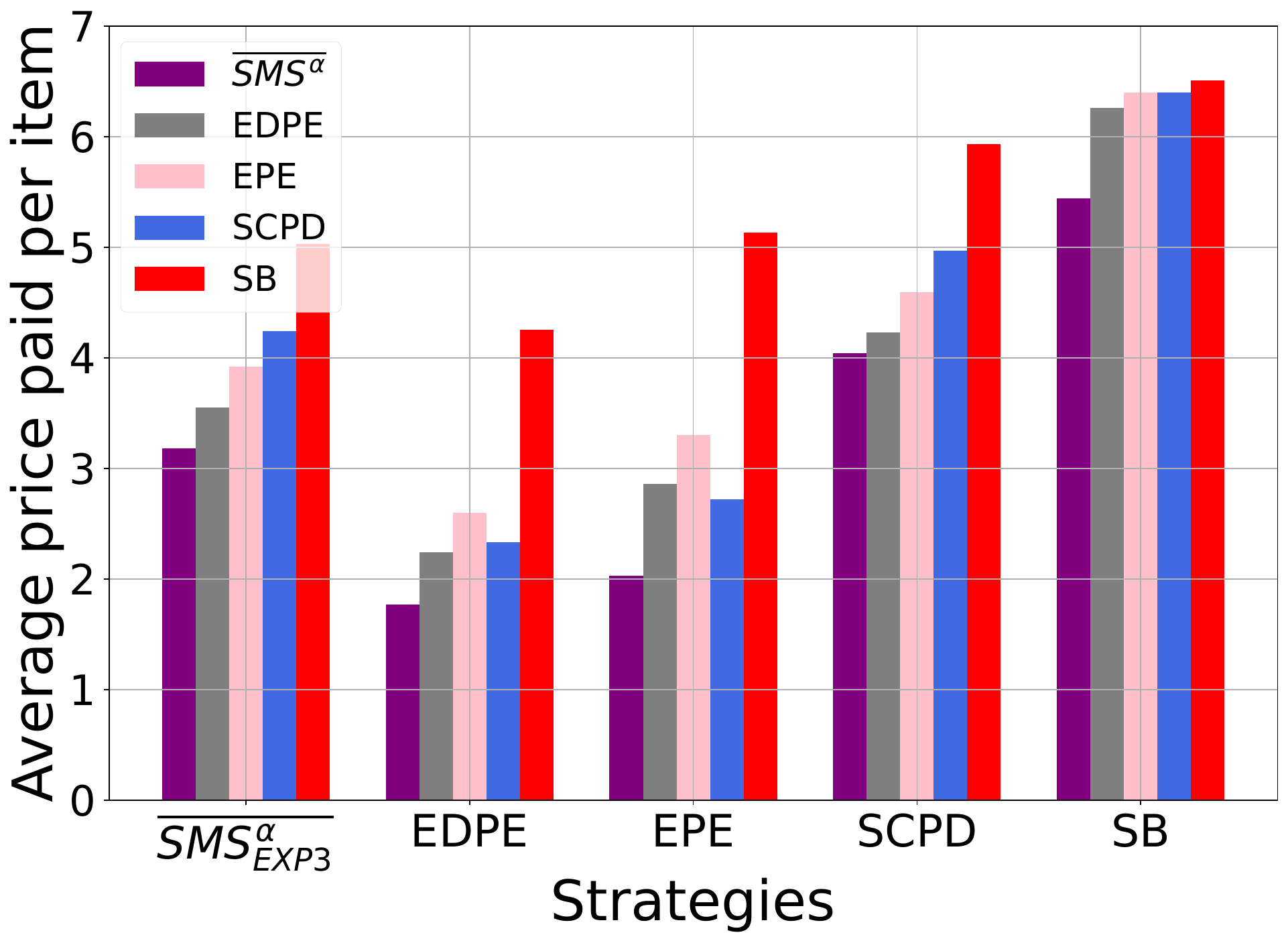}%
\label{avg price PP}}
\hfil
\subfloat[Ratio of items won]{\includegraphics[width=2.3in]{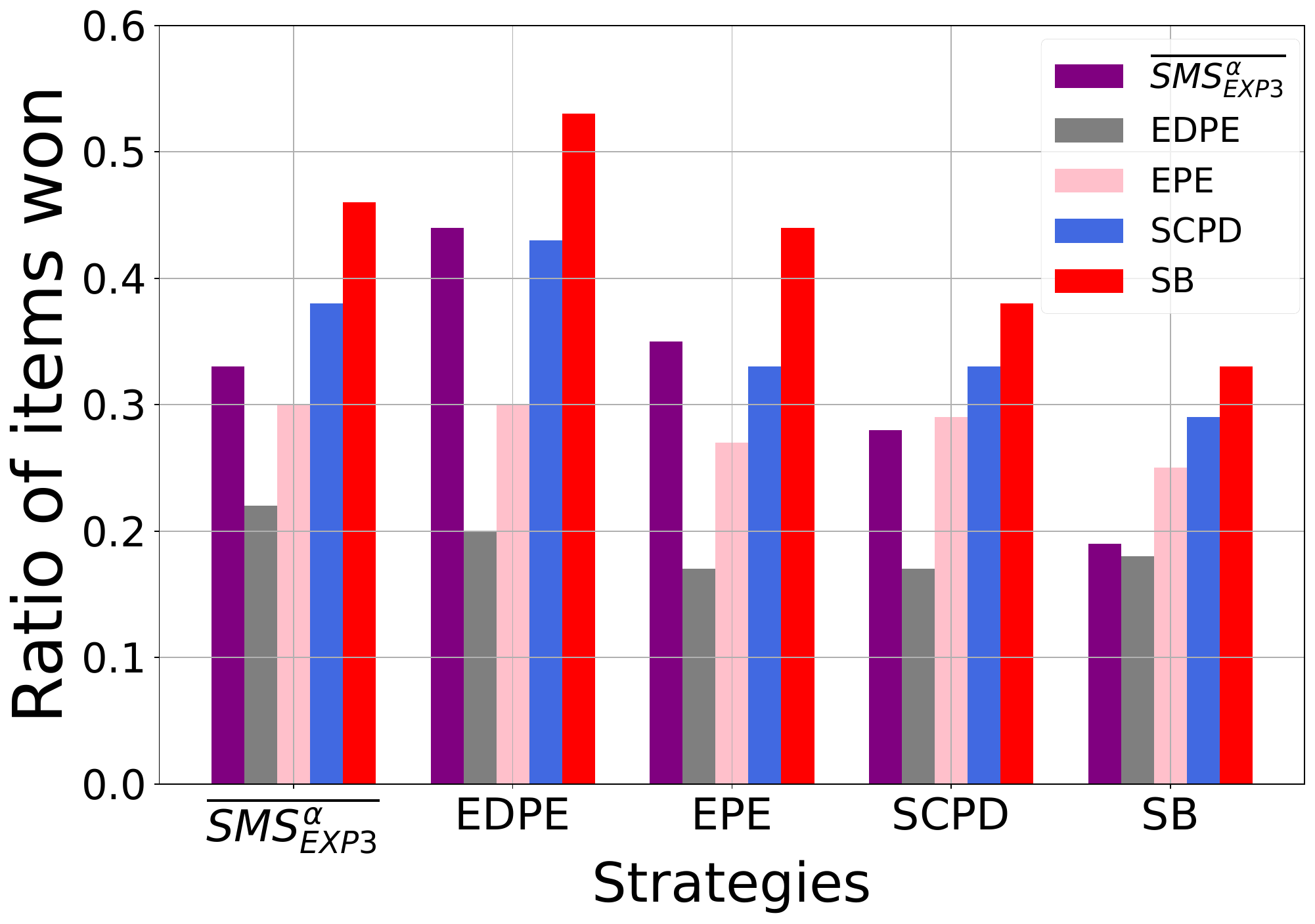}%
\label{ratio itm PP}}
\caption{Comparing own price effect of $\overline{SMS_{EXP3}^\alpha}$ with PPB approaches with a level of certainty $(\eta_v,\eta_b)=(0.5,0.5)$ through two performance indicators: the average price paid per item won and the ratio of items won}
\label{fig:Own_price_effect PP}
\end{figure*}

We illustrate here how $\overline{SMS_{EXP3}^\alpha}$ tackles the own price effect as it has the highest average price paid per item won against PPB approaches amongst our determinization approaches. We show in Figure~\ref{fig:Own_price_effect PP} the average price paid per item won and the ratio of items won by each strategy $A\in S_{PPB}\cup \{\overline{SMS_{EXP3}^\alpha}\}$ against every strategy $B$ on the x-axis. We see in Figure~\ref{avg price PP} that $\overline{SMS_{EXP3}^\alpha}$ acquires items at a lower price in average than any PPB strategy. Focusing on strategy profiles where all bidders play the same strategy, $\overline{SMS_{EXP3}^\alpha}$ obtains a lower average price paid per item won than EPE, SCPD and SB. Moreover, all items are allocated in the case of $\overline{SMS_{EXP3}^\alpha}$ whereas only $61\%$ of items are allocated when all bidders play EDPE, see Figure~\ref{ratio itm PP}. This last point mainly explains why the expected utility when all bidders play $\overline{SMS_{EXP3}^\alpha}$ is $25\%$ higher than if they all play EDPE. Similar results are observed for the other determinization approaches and different levels of certainty. Thus, our determinization approaches notably achieve greater coordination than PPB approaches by allocating all items at a relatively low price. 

When facing a strategy $A\in S_{SMS}$, $SDSMS_{EXP3}^\alpha$ obtains the highest average price paid per item won. In both cases, $DSMS_{EXP3}^\alpha$ always maintains the lowest average price paid per item won. Nevertheless, the relative difference in average price paid per item won between two different determinization approaches never exceeds $7\%$ regardless of the opposing strategy. Focusing on strategy profiles where all bidders play the same strategy, $DSMS_{EXP3}^\alpha$ pays $8.8\%$ and $15.2\%$ less per item won than respectively $\overline{SMS_{EXP3}^\alpha}$ and $SDSMS_{EXP3}^\alpha$. Moreover, when all bidders play the same determinization approach, all items are allocated.To sum up, our determinization approaches notably achieve greater coordination than PPB approaches by allocating all items at a relatively low price. Among the proposed solutions, $DSMS_{EXP3}^\alpha$ achieves the best performance when all players play the same strategy. 

\begin{figure}[h!]
\centering
\includegraphics[width=0.7\linewidth]{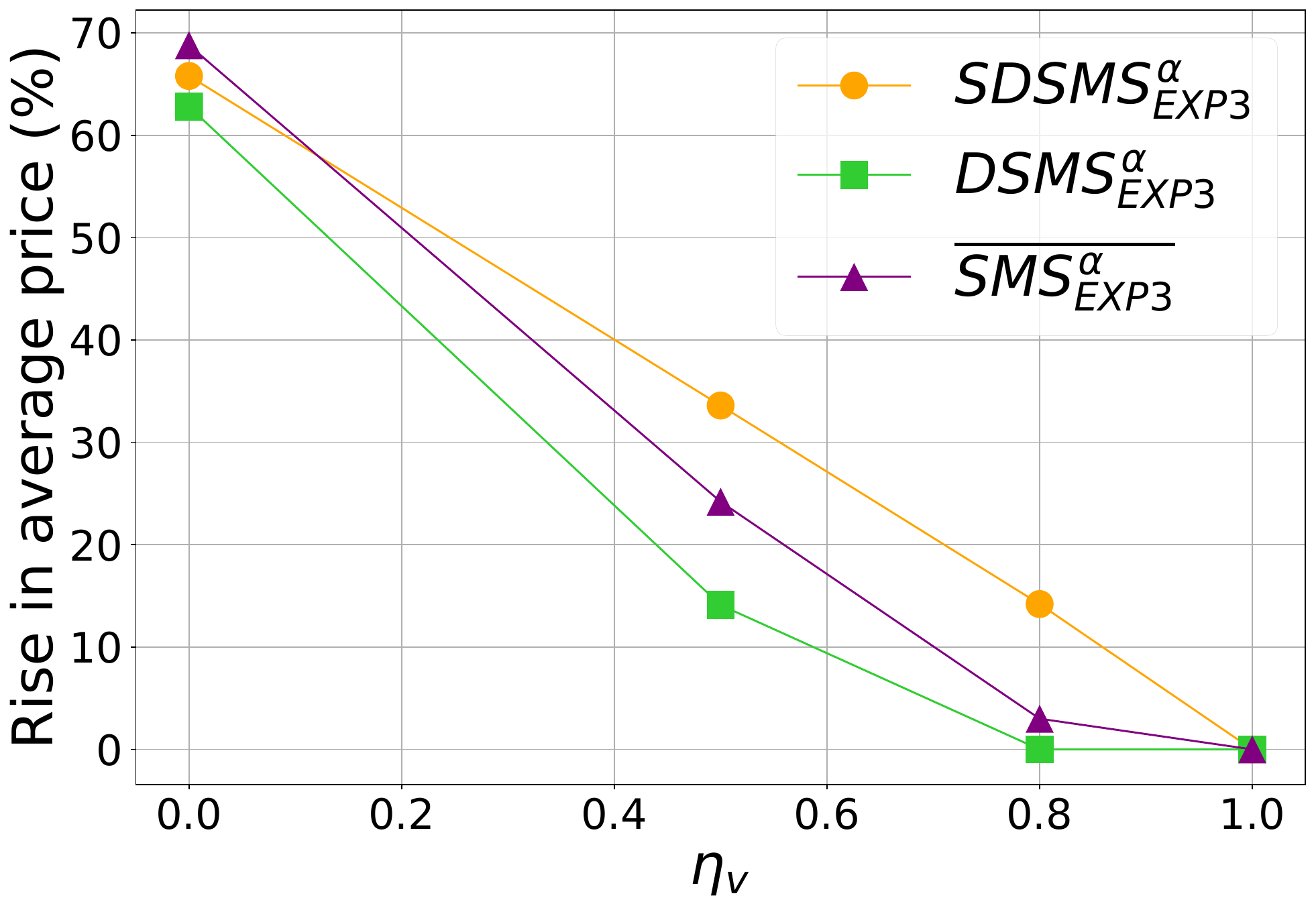}
\caption{Comparing the rise in average price paid per item won ($\%$) when all players play $CSMS_{EXP3}^\alpha$ and when all players play the same strategy $A$ with $A\in S_{SMS}$ with different levels of certainty ($\eta_v=\eta_b$).}\label{fig:Relative rise in avg price difference SAA-inc}
\end{figure}

Uncertainty significantly affects the coordination among bidders. In Figure~\ref{fig:Relative rise in avg price difference SAA-inc}, we show the rise in average price paid per item won when all bidders opt for the same determinization approach instead of $CSMS_{EXP3}^\alpha$. We observe that the average price increases with greater uncertainty for all approaches. Since all items are allocated in all cases, the rise in average price appears to be the primary cause of the decrease in expected utility with increasing uncertainty. 

\subsection{Lessons learned from the experiments}

Through these extensive experiments, we have seen that: 
\begin{itemize}
    \item Our three determinization approaches significantly outperform PPB approaches across all levels of certainty, notably by better tackling the own price effect and the exposure problem in budget and eligibility constrained environments.
    \item Differences between our determinization approaches are relatively small compared to the gains obtained against PPB strategies. This is true for all considered performance indicators and level of uncertainties. The strategy profile where all bidders play $SDSMS_{EXP3}^\alpha$ is a Nash equilibrium of the SAA-inc game with strategy set $S_{SMS}\cup S_{PPB}$. Moreover, $SDSMS_{EXP3}^\alpha$ reduces slightly more the risk of exposure than the other determinization approaches. However, in scenarios where all players adopt the same strategy, $DSMS_{EXP3}^\alpha$ better tackles the own price effect and yields higher expected utility. 
    \item Uncertainty impacts our three determinization approaches by increasing their average price paid per item won as well as their risk of exposure. Consequently, this diminishes their respective expected utility. Moreover, when all bidders are playing the same strategy, uncertainty significantly disrupts coordination among bidders, particularly in instances of wide type distributions.
\end{itemize}

Overall, these findings underscore the superior performance of our determinization approaches compared to PPB methods, while also highlighting the impact of uncertainty on their efficiency and coordination.

\section{Conclusions and Future Work} \label{sec:conclusion}

This paper introduces the three first bidding strategies that tackle simultaneously the \textit{exposure problem}, the \textit{own price effect}, \textit{budget constraints} and the \textit{eligibility management problem} in a simplified version of SAA with incomplete information (SAA-inc). Each strategy is computed through a different determinization approach that heavily relies on the SM-MCTS algorithm and a specific method for the prediction of closing prices. A simple inference method based on tracking bid exposure is used to update one's belief about opponents' budgets. These approaches employ a hyperparameter $\alpha$ that enables a bidder to arbitrate between expected profit and risk-aversion. We show that uncertainty highly impacts our determinization approaches even if they remain substantially better than state-of-the-art strategies, notably by obtaining higher profitability with less risk of exposure. Among our three determinization approaches, $SDSMS_{EXP3}^\alpha$ seems slightly more efficient than the two others. Indeed, the strategy profile where all bidders play $SDSMS_{EXP3}^\alpha$ constitutes a Nash equilibrium of the SAA-inc game for the set of considered strategies and playing $SDSMS_{EXP3}^\alpha$ leads to less exposure. However, better coordination between bidders playing the same strategy and hence better expected utility is obtained with $DSMS_{EXP3}^\alpha$.

Although we think that these determinization approaches can easily be improved by increasing the number of generated profiles or by modifying their selection phase with an even better selection strategy, we believe that the greatest area for improvement lies in inference methods in order to obtain narrower type distributions, before the auction but above all after each round during the auction.

\bibliographystyle{plain}
\bibliography{sample-base}

\end{document}